\documentclass[sigconf]{acmart}

\acmConference[ICPC 2022]{The 30th International Conference on Program Comprehension}{May 21–22, 2022}{Pittsburgh, PA, USA}

\usepackage{subfig}
\usepackage{graphicx}
\usepackage{xcolor}

\author{Rishab Sharma}
\affiliation{
    \institution{University of British Columbia}
    \country{Canada}
}
\email{rishab.sharma@alumni.ubc.ca}

\author{Fuxiang Chen}
\affiliation{
    \institution{University of British Columbia}
    \country{Canada}
}
\email{fuxiang.chen@ubc.ca}

\author{Fatemeh Fard}
\affiliation{
    \institution{University of British Columbia}
    \country{Canada}
}
\email{fatemeh.fard@ubc.ca}

\author{David Lo}
\affiliation{
    \institution{Singapore Management University}
    \country{Singapore}
}
\email{davidlo@smu.edu.sg}

\copyrightyear{2022}
\acmYear{2022}
\setcopyright{acmcopyright}\acmConference[ICPC '22]{30th International Conference on Program Comprehension}{May 16--17, 2022}{Virtual Event, USA}
\acmBooktitle{30th International Conference on Program Comprehension (ICPC '22), May 16--17, 2022, Virtual Event, USA}
\acmPrice{15.00}

\begin{document}
%
\title{An Exploratory Study on Code Attention in BERT}

\newcommand{\todoc}[2]{{\textcolor{#1} {\textbf{[[#2]]}}}}
\newcommand{\todoblue}[1]{\todoc{blue}{\textbf{[[#1]]}}}

\newcommand{\fuxiang}[1]{\todoblue{Fuxiang: #1}}



\begin{abstract}
Many recent models in software engineering introduced deep neural models based on the Transformer architecture or use transformer-based Pre-trained Language Models (PLM) trained on code. 
Although these models achieve the state of the arts results in many downstream tasks such as code summarization and bug detection, they are based on Transformer and PLM, which are mainly studied in the Natural Language Processing (NLP) field. 
The current studies rely on the reasoning and practices from NLP for these models in code, despite the differences between natural languages and programming languages. 
There is also limited literature on explaining how code is modeled. 

Here, we investigate the attention behavior of PLM on code and compare it with natural language. We pre-trained BERT, a Transformer based PLM, on code and explored what kind of information it learns, both semantic and syntactic. 
We run several experiments to analyze the attention values of code constructs on each other and what BERT learns in each layer. 
Our analyses show that BERT pays more attention to syntactic entities, specifically \textit{identifiers} and \textit{separators}, in contrast to the most attended token \texttt{[CLS]} in NLP. 
This observation motivated us to leverage \textit{identifiers} to represent the code sequence instead of the \texttt{[CLS]} token when used for code clone detection. Our results show that employing embeddings from identifiers increases the performance of BERT by 605\% and 4\% F1-score in its lower layers and the upper layers, respectively. When identifiers' embeddings are used in CodeBERT, a code-based PLM, the performance is improved by 21--24\% in the F1-score of clone detection.
The findings can benefit the research community by using code-specific representations instead of applying the common embeddings used in NLP, and open new directions for developing smaller models with similar performance.

\end{abstract}



\maketitle

\section{Introduction}
Pre-trained language models (PLMs) such as BERT  \cite{bert} and RoBERTa \cite{roberta} are based on Transformer networks \cite{all-you-need} and have been widely used in the field of Natural Language Processing (NLP). These language models are trained on a large corpus of data and are able to capture the general understanding of a language \cite{lmknowledge}. 
More recently, the software engineering community has employed language modeling on tasks such as code retrieval, program repair, code documentation, code clone detection, and bug detection \cite{scelmo, codebert, cubert, xia-trans}.
The results obtained by fine-tuning the language models pre-trained on code corpora outperformed the previous state-of-the-art models \cite{graphcodebert, codebert, codexglue}.
There have been attempts \cite{codebert, graphcodebert} to develop code-specific PLMs, which can leverage existing techniques in Software Engineering (SE) to enrich the underlying Transformer networks for SE-related tasks. 
However, all these works are based on studies in natural language and they do not investigate how code is modeled differently.

Under the hood, these transformer-based language models use the attention mechanism, which enables the model to attend to more important tokens in a sequence. This is used in multiple software engineering studies for developing models or explaining the models' predictions \cite{attend, code2seq, li, msr-context, shuai}.
The importance of this architecture with attention mechanism can be derived from the fact that employing the vanilla Transformer based networks without any additional techniques like utilizing abstract syntax tree and information retrieval can generate state-of-the-art results in various tasks such as code summarization, code retrieval, code clone detection, and program repair \cite{trans3, transformer, plbart, codebert, graphcodebert}. 
However, there is no study to explore how the PLMs attend toward different tokens in code and whether the assumptions made in NLP for representing a text sequence apply to code, 
despite the differences between programming languages and natural languages. For example, code comprises of different constructs such as identifiers and data-types, and how the PLMs model these constructs is unknown.

Although there are studies in software engineering \cite{difnlpl} that show the similarities and dissimilarities between the programming languages and natural languages, assuming similar behavior in code by directly applying the NLP knowledge may not be appropriate; and it is necessary to study how the Transformer architecture using language models learns the semantic and syntactic information in code. The literature on this topic is limited. It is only recently that Karmakar and Robbes \cite{new-ptm-study} studied the comprehension abilities of four different pre-trained models (BERT \cite{bert}, CodeBERTa \cite{codeberta}, CodeBERT \cite{codebert} and GraphCodeBERT \cite{graphcodebert}) trained on code. They compare the performance of these models from the perspective of probing classifiers without investigating the details of what the architecture learns about code.



This study evaluates the BERT model, trained on code corpus. BERT is chosen as it is the basis of RoBERTa used by CodeBERT and other PLMs developed for code. In addition, we want to compare our code-specific findings with what BERT learns about natural language in the study of Clark et al. \cite{whatdoes}. 
We evaluate the BERT model that we pre-train on code as a classical NLP pipeline. 
An NLP pipeline analyzes the language based on its linguistic features, which can contain semantic and structural properties \cite{nlppipeline1, nlppipeline2}. 
In our study, we explore the \textbf{attention behavior} of BERT on different syntactic types (i.e., code constructs). We want to understand \textbf{which layers} the model learns the semantic and syntactic knowledge of code. To compare our findings (BERT trained on programming languages) with (BERT trained on natural languages), a direct comparison is made with an existing NLP study by Clark et al. \cite{whatdoes}.
Our study uses BERT as it is a popular and strong baseline used in NLP and Software Engineering (SE) areas. Moreover, other variations of BERT such as RoBERTa \cite{roberta} are based on BERT and they have a similar architecture that is used in many PLMs on code \cite{codebert, graphcodebert}.
Our results show that, unlike NLP, the \texttt{[CLS]} and \texttt{[SEP]} tokens are not attended much in code.
This is important as the learned embedding of \texttt{[CLS]} and and \texttt{[SEP]} are frequently used in the classification tasks in both fields of NLP and SE.
However, we observed that the \texttt{[CLS]} is not the best token to capture the information of the code sequence for classification tasks. 
On the other hand, BERT attends more toward \textit{identifiers} (e.g., class names) and \textit{separators} (e.g., \texttt{;}, \texttt{\{}) in code. This led us to the hypothesis that the embeddings learned for identifiers in code can be a better representation in classification tasks related to code. 
To validate our hypothesis, we conducted experiments on code clone detection, where we compare the results obtained using the embeddings from \texttt{[CLS]} and \textit{identifiers}. 
When the identifier's embedding is used, the performance is improved by 605\% and 4\% in F1-score, using embeddings from the lower layers and the upper layers of BERT, respectively. 
We apply our findings on CodeBERT \cite{codebert}, a Transfomer based PLM developed for code representation, which uses a different pre-training objective. Similar behavior is observed when using the identifier's embedding, and the results of code clone detection are improved by 21--24\% in F1-score compared to when the \texttt{[CLS]} embedding is used. 

The contributions of this paper are listed as follows:
\begin{itemize}
  \item We present a first study that explores the \textbf{attention comprehension} abilities of BERT pre-trained on code. 
  \item We compare BERT's learning in the modeling of natural language \emph{English} and the programming language \emph{Java}. Our findings reveal some similarities and differences in modeling the languages by the PLMs.
  \item We study whether BERT can intrinsically learn code structure. Our analysis shows that the models can understand the general syntactic structure by predicting correct syntactic entities; however, they cannot correctly predict the valid tokens with high accuracy.
  \item  We propose to use the embeddings generated from identifiers in place of the vanilla embeddings used in BERT. When employed for code clone detection, these embeddings create better results than the vanilla techniques used in the PLM.

\end{itemize}

We have open-sourced our experiments for replication of the results and further usage in the software engineering community\footnote{\textcolor{black}{https://github.com/fardfh-lab/Code-Attention-BERT}}. 

\textbf{Significance of Study:}
Our study provides new insights 
on two important entities in code: \textit{identifiers} and \textit{separators}. This also includes how they are modeled in PLMs. The representation of these code constructs can be used to generate better models for code clone detection. The generated embeddings are effective as they increase the performance significantly using the lower layers of BERT. 
Researchers can use the findings presented in the paper to use the current PLMs and obtain better results. Moreover, though this is not the focus of our research, these findings can be {used to reduce model sizes}, addressing the lack of computational accessibility and enable a wider audience to use PLMs in practice, e.g. by having 6 layer models with similar performance to 12 layer models. 

The rest of this paper is organized as follows. 
In section \ref{background} we detail the necessary background. Study design and the experimental setup are presented in section \ref{strudyDesign}, followed by answers to each research question in sections \ref{rq1-results}, \ref{rq2-results}, and \ref{rq3-results}.  
Related works are discussed in section \ref{related-work}, followed by threats to validity in section \ref{threat}. We conclude the paper in section \ref{conclusion}.

\section{Transformer and BERT Architecture} \label{background}
In this section, we explain the necessary background. 

\subsection{Transformer}
Transformer forms the fundamental units of BERT, a popular PLM used in many studies \cite{bert}. A Transformer in a sequence generation setting uses multiple stacks of Encoder and Decoder  \cite{all-you-need}. Transformers primarily leverage Multi-Head Self Attention and Feed-forward neural network to model the relationship within the text \cite{all-you-need}. BERT solely uses the Transformer's encoder stack due to its pre-training objectives used in its pre-training \cite{bert}.

\textbf{Multi-Head Self Attention Layer}: The Transformer encoders use the self-attention \cite{all-you-need} mechanism to represent the input sequence by relating every input sequence element with each other. For the given input sequence S $(S=T_1, T_2, ... T_n  )$, of length $n$, each input token $T_i$ is first converted into its vector embedding $x_i$. 


Each encoder layer consists of three sets of weights matrices, $W(q)$, $W(k)$, and $W(v)$ that is used for query, key, and value, respectively. Initially, these matrices are randomly initialized and they are trainable parameters. The vector multiplication between input embedding $x_i$ and $W(q)$, $W(k)$, and $W(v)$ matrices generates unique query $q(i)$, key $k(i)$, and value $v(i)$ vectors, respectively.
\begin{equation}
Attention(Q,K,V) = \frac{Softmax(QK^{T})}{\sqrt{
d_{k}}}V \label{eq:attention}
\end{equation}

The attention scores are obtained by performing dot product between vector $q(i)$ and vector $k(i)$. This score is then divided by the square root of the dimension of the input embedding $d_{k}$. The softmax function is applied to the attention scores to obtain attention distribution as shown in \eqref{eq:attention}. Each softmax score is multiplied with the value vector $v(i)$ to generate weighted vectors added to get the final vector $z(i)$. This same task is repeated for each token multiple times, to form multi-headed attention. Each attention head learns a different vector space which is then combined by concatenating the information learned by each attention head.

\textbf{Position Embeddings}: Languages are sequential, and changing the order of words may lead to a completely different meaning or an erroneous sentence, which does not have any congruous meaning. In Transformer, a complete sequence of text forms the input into the multi-head attention layer, which helps in promoting parallelization. To maintain and learn the syntactic correctness of the language, an additional type of embeddings, position embeddings, is introduced. 
Specifically, the word embedding $x_i$ and its corresponding position embeddings $p_i$ are added, $xp_{i} =  x_{i} + p_{i}$. Embedding $xp_{i}$ is then further fed into the multi-head attention layer:

\textbf{Residual Connection and Add Normalize}: The output of multi-head attention is added to the residual connection from $x(i)$, and layer normalization is performed on the added vector. 

\textbf{Feed Forward Neural Network Layer}: The output from the Add Normalize layer is then fed into the feed-forward neural network, which has a common architecture for all the tokens. However, this step is done in parallel on each token as shown in equation \eqref{eq:ffn}.
\begin{equation}
FFN(x_{i}) = max(0, x_{i}W_{1} + b_{1})W_{2} + b_{2}\label{eq:ffn}
\end{equation}

The output is then added and normalized. This finally completes a single feed-forward propagation in one encoder unit. The output from this encoder unit is fed into the next unit, and the same processing is done on it.

\subsection{BERT: Bidirectional Encoder Representations From Transformers}

BERT is a PLM trained with the core motivation to build a bi-directional model, which can learn the general understanding of the language and then it can be fine-tuned for various downstream tasks.

\textbf{BERT-Pretraining}: BERT is pre-trained on two unsupervised tasks (i) Masked Language Modelling (MLM) and (ii) Next Sentence Prediction (NSP). Originally, BooksCorpus and English Wikipedia datasets were used to train BERT \cite{bert}. The datasets contain books and Wikipedia information that is available online.

\textbf{Masked Language Modeling (MLM)}: Unlike other neural language models, BERT is pre-trained using Masked Language Modeling. In masked language modeling, the task is to predict a randomly masked token in a sequence rather than predicting the next token in sequence. Within Masked Language Modeling in BERT, 15\% of the tokens are randomly masked. Masking of the tokens follows a probability pattern where a token is either replaced with (1) a [MASK] token 80\% of the time, (2) a random token 10\% of the time, or (3) it will remain unchanged for 10\% of the time.\\
\textbf{Next Sentence Prediction (NSP)}: BERT is also pre-trained on a binary next sentence prediction task. This task focuses on training a language model that can discern whether the given sentence is the next sentence of the current sentence. This is done by randomly replacing the next sentence 50\% of the time with random sentences. This would enable the model to discern the difference between the natural coherence of the sentences, which motivates the model to learn the general understanding of a language.

\textcolor{black}{\textbf{Special Tokens in BERT}: BERT introduces special tokens namely \texttt{[CLS]} and \texttt{[SEP]} tokens as its input formatting. \texttt{[CLS]} is appended at the start of the sentence, which is supposed to learn knowledge of the whole sentence. \texttt{[SEP]} token is used as a ``separator'' between sentences to highlight the end of a sentence. The representation from the \texttt{[CLS]} token is also used  during the NSP pre-training task of the BERT model, as a ``classification'' token to predict whether a sentence is the next sentence of a previous sentence.}

\section{BERT Attention for Code} \label{strudyDesign}
The Transformer-based model primarily uses the multi-headed attention to model languages \cite{bert, roberta}. Therefore, attention plays a vital role in training the PLMs, and thus it is important to study the attention behavior of PLMs on the given text information. An attention distribution for a given sentence can quantify the Transformer model's importance to a particular token. A higher attention value on a specific token means that token has a higher weight in determining the representation of a given sentence, making it a significant token to learn and represent a sentence. Therefore, knowing the important tokens within the code is important for extracting meaningful information from the PLMs. To understand and visualize this attention information, we plot the attention maps for different kinds of token types, similar to a study released for the English language by Clark et al. \cite{whatdoes}.
We use this study as an initial comparison of Transformer learning between code and natural language, and design different experiments to understand the behavior of this architecture for programming languages.

\subsection{Experimental Setup}
There are two variants of BERT: BERT-Large and BERT-Base \cite{bert}. BERT-Base model contains 12 stacked encoders, with 12 attention heads in each layer. 
In our study, we use the BERT-Base cased language model as a direct comparison against the natural language, English. The cased language model is used so that the BERT model can learn the Java coding conventions for the identifiers, which promotes the use of different casing styles.
As we intend to compare our results of the BERT model trained on code with natural language, we need to keep the models the same. 
There are multiple implementations available for BERT and its variants trained on code, which are pre-trained with different objectives (e.g., MLM and Replaced Token Detection). 

However, for the comparison, we required a BERT model that is pre-trained with the same objectives as the one used for the NLP study \cite{whatdoes} and also should purely be pre-trained on code. 
We are unable to find an open-source pre-trained BERT-Base model that is trained solely on code.
Therefore, we trained a BERT-Base cased model from scratch. All the pre-training decisions like the number of layers in the encoder and decoder, the casing of tokens, and the pre-training objectives are consistent with the experimental settings in the NLP study by Clark et al. \cite{whatdoes}. For pre-training the BERT model, we used the benchmark dataset released by Husain et al. \cite{codesearchnet}. This dataset contains a large corpus for six different programming languages collected from open source repositories available on GitHub. However, we only confine our work to the Java programming language in this study (we do not require a multilingual model) to keep it consistent with the NLP study where the BERT model is only trained on a single natural language. Java is chosen as the language in our study because it is a widely adopted programming language to study program comprehension in the software engineering community. It is also a commonly used programming language in the industry.
The statistics of the dataset used is shown in Table \ref{table: bert-dataset}.

\begin{table}[!htb]
\centering
\caption{Dataset details for pre-training BERT}
\label{table: bert-dataset}
\small
\begin{tabular}{cccl}

\hline
\textbf{Split} &\vtop{\hbox{\textbf{\# Records in the Dataset}}} \\
\hline

Train & 164,923 \\
\hline

Validation & 5,183\\
\hline

Test & 10,955\\
\hline

\end{tabular}
\end{table}

For training the BERT-Base cased model, we used the open-source implementation of BERT that is available from its authors, Devlin et al. \cite{bert}. 
In this study, as we focus only on code representation using BERT, we remove the JavaDoc comments, the inline comments, and the block comments from the dataset. This cleaned dataset only contains code and it is used for training the BERT model.
Further, to make the dataset eligible for the next sentence prediction task, we keep the separation of the code sentences in different lines, as mentioned on the BERT's official implementation repository\footnote{https://github.com/google-research/bert} and this separation process is also used in previous study \cite{ase-ptm}.
We then train BERT using this dataset. The training continues for four consecutive days on a Tesla V100 GPU with 32GB memory. The performance of the BERT model on the validation dataset for the two pre-training objectives, Masked Language Modeling and Next Sentence Prediction, is shown in Table \ref{table: bert-perf}. The results obtained for BERT's pre-training for the NSP objective on code is 94.75\%, which is close to the original pre-training result for BERT model on English, as reported by the authors of the BERT model. 
This result confirms that the model we have trained is consistent with the original BERT model on English language.
The accuracy achieved by BERT for MLM is not reported in the original paper. Therefore, we cannot make any direct comparison for this pre-training objective. Nonetheless, we believe that 87.44\% is a reasonable accuracy for predicting the masked tokens, as it involves the correct prediction of the code token in a sentence.

In the remaining paper, the BERT model that we refer to is the model that we have pre-trained on Java.

\begin{table}[!htb]
\centering
\caption{Accuracy of pre-trained BERT-Base on Valid dataset. The results are close to the original pre-training results reported in BERT}
\label{table: bert-perf}
\small
\begin{tabular}{cccl}
\hline
\textbf{Pre-Training Task} &\vtop{\hbox{\textbf{\% Accuracy}}} \\
\hline
Masked Language Modeling & 87.44 \\
\hline
Next Sentence Prediction & 94.75 \\
\hline
\end{tabular}
\end{table}

\subsection{Research Questions}

This study investigates answers to the following research questions, which require analysis of the BERT's attention behavior and comprehension abilities for code. 

\textbf{RQ1 (Attention Behavior): What is the surface level behavior of BERT attention over code?} In this research question, we study the attention scores set by BERT on the special tokens available within BERT. Moreover, we also study the divergence of information learned by different attention heads in various layers of BERT. The relative position on tokens is used to investigate the short or long-range attention over different layers.
    
\textbf{RQ2 (Code Construct Relationships): How does BERT encode the code-specific relationships among different code constructs?} In this research question, we explore the code constructs that receive the highest attention, which can reveal important code constructs learned by the model. Furthermore, we investigate how each code construct, specifically identifiers, distributes its relationship with other code artifacts like separators and keywords in different layers. Moreover, we study whether BERT can encode syntactic information of code by using a non-parametric probing.
    
\textbf{RQ3 (Semantic Knowledge): How does BERT learn the semantic knowledge in code?} In this research question, we investigate which layers within the BERT model can learn the semantic understanding of code. We do this by analyzing the performance of each layer for the code clone detection task. Finally, based on our findings from RQ1 and RQ2, we leverage the embeddings learned from identifiers to better represent code in place of the vanilla \texttt{[CLS]} token.

\subsection{Generating Attention Maps}

Once BERT is pre-trained on the Java dataset, we use the model's trained weights to extract the attention distribution. 
For extracting the attention scores, following Clark et al. \cite{whatdoes}, we do not use the train dataset for understanding the attention mechanism. Instead, we select samples from the test dataset. 
As BERT uses sentence piece tokenization of text, it can lead to sub-tokenization of a token into multiple tokens depending on the vocabulary used within the tokenizer. Therefore, we can sample code examples using two different strategies (i) Select only code examples from the test dataset for which the length before and after the tokenization remains the same, (ii) Randomly sample code where tokens are sub-tokenized and are later combined for attention analysis. We conducted the same experiments using both strategies; however, the results are similar irrespective of the sampling strategy. Therefore, to avoid redundancy, we only include the experiments of the second strategy and combine the tokens for the analysis.

\section{Attention Behavior} \label{rq1-results}
 Here, we would only focus on the surface-level behavior of BERT's attention on code. We follow the surface level analysis as conducted in \cite{whatdoes}, and only include the attention over special tokens, the relative positioning of the tokens, and the similarity and difference in attention distribution over different layers and heads.
 
\subsection{Attention on Special Tokens}\label{special-tokens}
Before studying the attention behavior of code constructs, we look at the attention behavior over the special tokens introduced by the BERT's pre-processing. It is important to study these special tokens as they play an important role in pre-training the BERT model.
Two special tokens are introduced to feed input to the BERT model, namely \texttt{[CLS]} and \texttt{SEP}. The \texttt{[CLS]} token's representation is used during pre-training for the next sentence prediction objective and assumed to contain the complete information about the sequence, and the \texttt{[SEP]} token acts as a delimiter between two sentences.

\begin{figure}[!htbp]
    \includegraphics[scale=0.5]{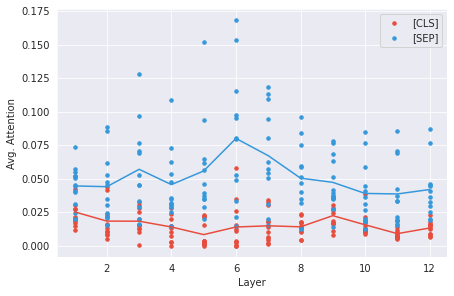}
    \caption{Attention on special tokens \texttt{[CLS]} and \texttt{[SEP]}}
    \label{fig:attcls}
\end{figure}

To visualize the attention distribution of these tokens, we apply the following process. The BERT architecture we use has 12 layers, and each layer contains 12 attention heads. As discussed previously, each attention head is trained to learn different aspects of the input, and when all the heads are combined, they should reflect the complete knowledge. Therefore, for each layer and each head in the layer, we aggregate the attention scores of the model for the \texttt{[CLS]} and \texttt{[SEP]} tokens, separately. We aggregate the total attention scores on these special tokens (the scores for each of the \texttt{[CLS]} and \texttt{[SEP]} are computed separately), which is the attention given to these tokens by all the other words in the sequence. This score is then normalized with the token's total number of occurrences, giving us the average attention put on each of \texttt{[CLS]} and \texttt{[SEP]} tokens in each given code. The results are shown in Figure \ref{fig:attcls}. The multi-scatter plot on the y-axis of each layer shows the values of different attention heads at each layer.

As shown in Figure \ref{fig:attcls}, both of these special tokens have very low attention values, which means BERT has little focus on the \texttt{[CLS]} and \texttt{[SEP]} tokens when modeling code.
This reduces the significance of special tokens in learning the relationship of code tokens. This result is contrary to what has been seen in the NLP study \cite{whatdoes}. In NLP, the average attention on \texttt{[CLS]} and \texttt{[SEP]} ranges between 0 to 0.6 over different layers. This score ranges only between 0 to 0.10 for code.

For the \texttt{[SEP]} token, the average attention increases between Layers-4 to Layers-6, and then there is a consistent decline until layer 10, and then remains uniform for the remaining layers. Overall, the \texttt{[SEP]} token consistently gets more attention over the \texttt{[CLS]} token in all layers. However, in the case of NLP, Clark et al. \cite{whatdoes} reported that the initial layers until Layer-4 attend more towards \texttt{[CLS]} over the \texttt{[SEP]} token, which is not the case as seen in Figure \ref{fig:attcls}. We see a gradual decrease for the \texttt{[CLS]} token until Layer-5 and then a gradual increase until Layer-9. 

\subsection{Attention on Relative Positioning of Tokens}\label{context}
\begin{figure}[!htbp]
    \includegraphics[scale=0.5]{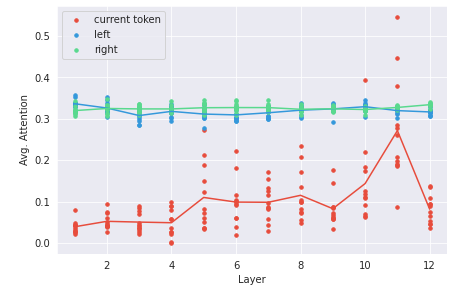}
    \caption{Attention on relative position of tokens}
    \label{fig:relative_position}
\end{figure}

The attention score of the relative positioning of the tokens can reveal the range (long and short) of attention distribution learned by the model. Attention towards immediate neighbor tokens means the model primarily understands the local context of the code. In this study, we find the model's attention that is put on the current token, previous token, and the next token. As shown in Figure \ref{fig:relative_position}, BERT focuses less towards the token itself while learning its representation when compared to attention towards its left and right token. In the initial layers, the concentration on the token itself is lower, which gradually increases in the middle layers (Layer-5 to Layer-9). At Layer-11, the attention of the current token increases and it is close to the attention value of its left and right token. 

On average, 30\% of attention is put by BERT on the left and right token. Uniform attention values are seen over each layer. Also, it demonstrates that the BERT for code model attends better locally than the long-range attention. This is similar to what we see in NLP, as BERT attends more towards local tokens in the English language.

\subsection{Attention Redundancy at Attention Heads and Layers}
BERT uses a multi-head self-attention mechanism to learn the representation of tokens. Multiple attention heads within the same encoder layer are expected to learn different language features. The variety of information learned by these attention heads is then concatenated to generate the representation for the language. Hence, it is important to study how much knowledge these attention heads learn from each other.

\begin{figure}[!htbp]
    \includegraphics[scale=0.5]{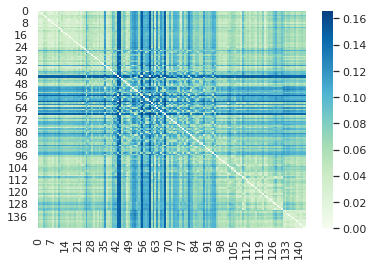}
    \caption{Attention redundancy at attention heads and layers}
    \label{fig:attduplicate}
\end{figure}

To understand this information, the Jensen-Shannon divergence \cite{jsd} between different attention heads in the BERT model is calculated. Jensen-Shannon divergence finds the similarity between two probability distributions. Therefore, we use the probability distribution of these attention heads to learn the differences. The more the distance between the two heads, the more disparate information is retained by them. In total, BERT contains 144 attention heads. Therefore, we have a $144\times144$ matrix to represent such information, as shown in Figure \ref{fig:attduplicate}.  
The middle layers retain diverse information compared to the lower and higher layers. Similar to the NLP study, redundancy in the attention distribution insinuates that not all attention heads learn unique information. Therefore, there is scope for applying model distillation strategies that can help reduce the model size and maintain equivalent performance. The only difference we see between Java and English is the absolute divergence values. The divergence in the information learned in different attention heads in English is slightly more than the  code, possibly due to the repetition of code tokens and code syntax \cite{difnlpl}.

\textbf{RQ1 Summary:} We studied the surface level behavior of BERT for code. Our experiments show that this behavior for code is mostly similar to NLP. Therefore, surface-level behavior is dependent more on the architecture of BERT, and the language differences have a minor effect on the surface-level behavior. The primary finding is the difference between code and NLP in the amount of attention learned by BERT for the special tokens \texttt{[CLS]} and \texttt{[SEP]}, which is much less for code; indicating that the \texttt{[CLS]} representation cannot efficiently capture the information about the code sequence. 

\begin{figure}[!htbp]
    \includegraphics[scale=0.5]{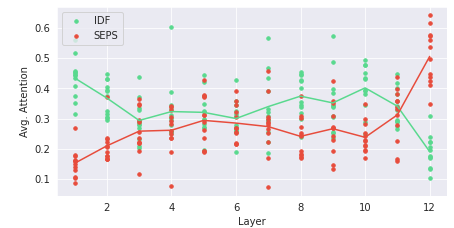}
    \caption{Attention on \texttt{identifier (IDF)} and separators SEPS}
    \label{fig:attidf}
\end{figure}

\section{Code Construct Relationships}\label{rq2-results}
Code has rich grammar and syntax. It contains multiple syntactic constructs such as \textit{identifiers}, \textit{separators}, and \textit{data types}. These code constructs are not available in natural languages. Therefore, we study the comprehension abilities of BERT for these syntactic types and the relationship of each syntactic entity with other code constructs. 

\subsection{Analysing Attention on Code Constructs in Code}\label{token-study}

Syntax plays an important role to increase the performance of neural models on different software engineering tasks, including code comment generation \cite{clair, hu1, rencos}, code clone detection \cite{graphcodebert, codebert}, program repair \cite{cv3}, software vulnerability detection \cite{cv2}, type prediction \cite{type-pred}, and bug detection \cite{cv5}. Here, we study the attention scores of BERT on the code constructs.

We draw an attention distribution plot similar to the plot for special tokens, \texttt{[SEP]} and \texttt{[CLS]} as done in section \ref{special-tokens}. In this plot, we collect the attention distribution for each of the syntactic types (i.e., code constructs). In order to tag each token to a syntactic type, we use the JavaLang Parser\footnote{https://github.com/c2nes/javalang}. The JavaLang parser provides both parser and tokenizer for the Java programming language and it has been used widely in other software engineering studies \cite{ase-ptm, rencos, ase}. The syntactic types included by the parser are \textit{identifiers}, \textit{separators} (e.g., `\texttt{,}', `\texttt{;}', `\texttt{\{}', `\texttt{\}}'), \textit{operators} (e.g., \texttt{=}, \texttt{+}, \texttt{-}), \textit{data-types} (\texttt{int}, \texttt{float}, \texttt{double}, etc.), \textit{keywords} (e.g., \texttt{abstract}, \texttt{continue}, \texttt{const}), and \textit{access-modifiers} (e.g., \texttt{public}, \texttt{private}). We group all the \textit{separators} into a single token \texttt{SEPS}. Note that the \texttt{[SEP]} token in Figure \ref{fig:attcls} is different from \texttt{SEPS} in Figure \ref{fig:attidf}. The former is a special token introduced by BERT to act as a delimiter between two sentences, and the latter is for the syntactic types (e.g., `\texttt{,}', `\texttt{;}', `\texttt{\{}') available within code. 

\begin{figure}[!htbp]
    \includegraphics[scale=0.5]{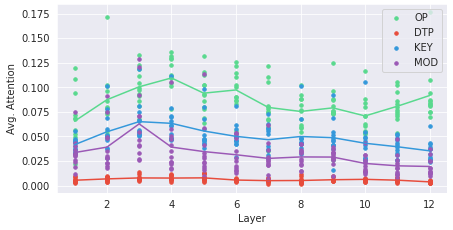}
    \caption{Attention on Syntactic types including Operators (\texttt{OP}), Data-type (\texttt{DTP}), Keywords (\texttt{KEY}) and Modifier (\texttt{MOD})}
    \label{fig:attother}
\end{figure}

The plots are shown in Figures \ref{fig:attidf} and \ref{fig:attother}. Figure \ref{fig:attidf} demonstrates that identifiers \texttt{(IDF)} and separators \texttt{(SEPS)} get the most amount of attention. The higher attention values over {identifiers} and {separators} demonstrate that BERT considers them as important tokens for code modeling. 
Therefore, both of these tokens should encode the latent knowledge of code better than the \texttt{[CLS]} and \texttt{[SEP]} special tokens. 

Interestingly, BERT models most of its attention over {identifiers} inherently, which could be a possible reason for the success of the BERT model. 
Between {identifiers} and {separators}, {identifiers} have higher attention values ranging between 30\%-45\%, until Layer-11. At the final layer, the {separators} tend to get the most amount of attention. A possible reason for high attention values for {identifiers} and {separators} could be that both types are the most frequently occurring tokens in the code. Also, the same token repeats at different positions with the same code sample due to the structure of the code \cite{difnlpl}.

{Literature on program comprehension has also acknowledged the importance of identifiers for easier program understanding \cite{idf1, idf2, idf3, pos}. 
Based on the obtained results of Figure \ref{fig:attidf}, a question is that \textit{``when modeling code, can using the \texttt{IDF} embeddings boost the performance of the same model over using the embeddings from \texttt{[CLS]''}? } We answer this question in section \ref{IDF-experiment}.
}

The attention plots of other syntactic tokens, including operator \texttt{(OP)}, data-type \texttt{(DTP)}, keywords\texttt{ (KEY)}, and access-modifiers\texttt{ (MOD)} are shown separately in Figure \ref{fig:attother}.
They are shown separately in Figure \ref{fig:attother}, as their attention scores are very low compared to identifiers and separators. 
Data type gets the least attention, followed by modifier, keyword, and operator. Such low attention values for these code constructs are surprising because Java is a statically typed language, and knowing data types is important for the correct compilation of code. Moreover, all of these syntactic types are related to the structural part of the code.

\subsection{Modeling the Relationship Among Syntactic Tokens and Identifiers}

\textcolor{black}{The results shown in the previous section indicate that BERT has the most attention on identifiers. As identifiers have previously shown to be important code constructs for program comprehension \cite{idf1, idf2, idf3, pos}, we analyze where identifiers focus when BERT models the identifiers' representation.}

\begin{figure}[!htbp]
    \includegraphics[scale=0.5]{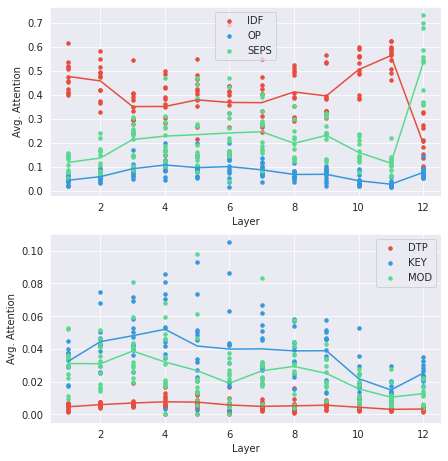}
    \caption{Modeling the relationship among syntactic tokens with identifier}
    \label{fig:idf_all}
\end{figure}
To analyse this attention distribution, we extracted the attention matrix of indexes where identifiers were available in code. Then, we leverage the identifier attention distribution, to model its relationship with other identifiers (\texttt{IDF}), separators (\texttt{SEPS}),operators (\texttt{OP}), data type (\texttt{DTP}), keywords (\texttt{KEY}) and modifiers (\texttt{MOD}) as shown in Figure \ref{fig:idf_all}.

\begin{figure}[!htbp]
    \includegraphics[scale=0.35]{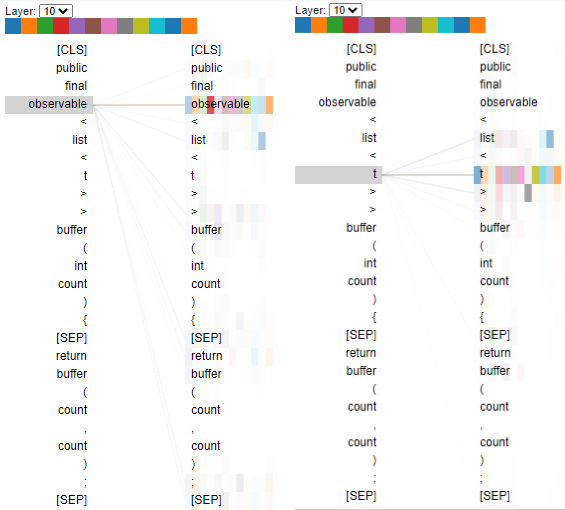}
    \caption{Attention of identifiers \texttt{observable} and \texttt{t} at layer 10. This representation has been generated using Bertviz \cite{bertviz}}
    \label{fig:learn_itself}
\end{figure}

As seen in the top plot of Figure \ref{fig:idf_all}, from Layer-1 to Layer-11 identifiers attends most towards identifiers, similar to the overall attention behavior of the BERT model. This is followed by separators and operators within the code. Again, data-type, keyword, and modifiers have the least attention values. As discussed in section \ref{context}, the attention is generally focused on local context. Hence, this attention behavior of identifiers should also be from the local context. 

Figure \ref{fig:learn_itself} shows an example of a code sequence and the attention distribution of tokens. There are two identifiers \texttt{observable} and \texttt{t}, for which the attention distribution over other tokens is highlighted, with links to those tokens. The darker the link, the higher is the attention value towards that token. We observe that these identifiers attend mostly on themselves. Each of them models its relationship with other local tokens (i.e., its immediate left and right tokens) to learn their representation, focusing on longer distance tokens with low attention values. Identifiers are generally associated with a data type; however, BERT fails to understand such associations intrinsically as data-type gets the least attention.

\subsection{Predicting Syntactic Entities Within Code}

To further investigate the abilities of BERT to capture the syntactic information of code, we conducted an experiment to understand whether BERT can generate the correct syntax of Java programming language that is compiled using a compiler. Please note that here we do not want to find the accurate token; rather, we aim to know whether BERT can accurately predict the syntax type of the missing tokens that follow the Java language's syntactic rules. 
We use a non-parametric probing task as used in \cite{codebert}. In this task, we mask the tokens available in the code. However, we mask these tokens based on their syntax type. For instance, given a code sample, we first mask all the \textit{identifiers} within the code and test the ability of the BERT model in predicting the identifiers in the masked spaces. Similarly, we repeat this masking for \textit{data-type}, \textit{decimal integer}, \textit{keywords}, \textit{modifier}, \textit{operator} and \textit{separators}. The results are shown in Table \ref{table:structure}.

\begin{table}[!htb]
\centering
\caption{Predictions for Syntactic Entities}
\label{table:structure}
\small
\begin{tabular}{ccccccl}
\hline
\textbf{Type} & \textbf{Precision} & \textbf{Recall} & \textbf{F1} \\
\hline
data type & 0.66 & 0.96 &0.78\\
\hline

decimalinteger & 0.59 &0.69 &0.61\\
\hline

identifier & 0.98 &0.88 &0.93\\
\hline

keyword & 0.68 &0.95 &0.79\\
\hline

modifier & 1.00 &0.99 &0.99\\
\hline

operator & 0.93 &0.98 &0.96\\
\hline

separator & 0.98 &0.97 &0.98\\
\hline

\end{tabular}
\end{table}

We evaluate the performance of this task using Precision, Recall, and F1 scores as follows: 
  Precision =  {TP}/{(TP + FP)}\label{precision}
;
  Recall =  {TP}/{(TP + FN)}\label{recall}
;
F1 =  2$\times${Precision$\times$Recall}/{(Precision + Recall)} \label{f1}.
Here, True Positive (TP) is the number of tokens correctly identified to replace the MASK tokens. 
False Positive (FP) is the number of samples incorrectly labeled as belonging to a group.
False Negative (FN) is the sample model incorrectly marked as not belonging to a group.

Overall, BERT can correctly predict token types for identifier, separator, modifier, and operators.
Syntactic types decimal integer, data-type, and keyword have lower performance than other types. The lower scores of these types can be related to the lower attention values they receive during the modeling as described in Section \ref{token-study}. The obtained scores for modifiers are high, despite the lower attention values. This can be related to the fact that the total number of modifiers in Java is less than five, making it an easy prediction.

{\textbf{RQ2 Summary:} 
BERT can learn different code constructs with varying performance. Mostly, BERT focuses on \textit{identifiers} and \textit{separators}. Therefore, the representation of identifiers can be useful for the downstream tasks. Also, the representation learned for an identifier is mostly contributed by the token itself or its immediate left and right tokens. }

\section{Semantic Knowledge} \label{rq3-results}

Semantic learning is pivotal to any language. It helps the model learn representation  meanings/ relationships within the learned space. 
In this section, we use code clone detection, a classification task that requires semantic information of code, to learn semantic information captured by BERT in each layer.
\textcolor{black}{Code clones is chosen as it provides the ability to study both the semantic and structural properties of code using one task.}
In this experiment, we use the embedding of the code sequence at each layer of BERT for identifying code clones. Code clone detection is chosen as it requires the model to capture some semantics about the code, where code samples can have structural or semantic equivalences.

This experiment is two fold: First, we assess the ability of BERT in capturing the semantics of code in different layers, using the learned embedding of the \texttt{[CLS]} token. In the second part, we seek the answer to our question: whether we can improve the performance by using the embeddings from \texttt{IDF} instead of \texttt{[CLS]}, based on our findings in RQ1-2, which are explained below.

\subsection{Semantic Representation of Code for Code Clone Detection Using \texttt{[CLS]}} \label{CLS-CCD}

Code clones are code samples that are identical to each other \cite{clone1, clone2}. These code samples can have structural or semantic equivalences \cite{bigclone}. Generally, code clone detection finds target code that has similar attributes to the given code \cite{bigclone}. If the code samples have the same functionality, they are tagged as semantic clones. Otherwise, they are classified as non-clones of each other. For the code clone detection, we use the pipeline made available by Lu et al. \cite{codexglue}, with few changes. The original pipeline was built for ROBERTa \cite{roberta}, which is changed to accommodate the BERT model. The pipeline uses the BERT model to initialize the encoder, and then a linear head is used over the pre-trained model for the classification into clones and non-clones. \textcolor{black}{The classification model is trained end-to-end, and, similar to works in NLP \cite{nlpclass, nlppipeline2}, we freeze the weights of the BERT model because we want to learn the model's capabilities on the target task without further training the BERT model.
Therefore, the weights of the encoder are \textbf{not} fine-tuned on the task, and we can only extract the embeddings to assess the semantic \textcolor{black}{capabilities} of the pre-trained model. \textcolor{black}{The linear head over BERT encoder is still fine-tuned on the task of code clone detection. The linear head is not originally part of the BERT model and it is only introduced as a projection for BERT representation. The linear head uses the hidden representation from BERT to detect the code clones}. The fine-tuning is done for three epochs, following similarly to how fine-tuning was done in the original BERT paper \cite{bert}. For fine-tuning, we use the Big-Clone-Bench dataset \cite{bigclone}. The statistics of the dataset is shown in Table \ref{table:clonedataset}.}

\begin{table}[!htb]
\centering
\caption{Dataset details for Code Clone Detection}
\label{table:clonedataset}
\small
\begin{tabular}{cccl}

\hline
\textbf{Split} &\vtop{\hbox{\textbf{\# Records in the Dataset}}} \\
\hline

Train & 901,028 \\
\hline

Validation & 415,416\\
\hline

Test & 415,416\\
\hline

\end{tabular}

\end{table}

\begin{figure}[!htbp]
\center
    \includegraphics[scale=0.42]{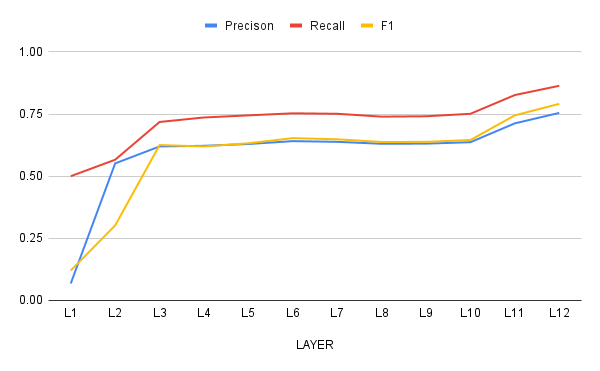}
    \caption{Performance of BERT for Code Clone Detection with \texttt{[CLS]} token embedding}
    \label{fig:clone}
\end{figure}

\begin{table}[htb]
\centering
\caption{Using BERT \textit{identifier} embedding for code clone detection}
\label{table:idf-clone}
\scalebox{0.85}{
\small
\begin{tabular}{ccccccl}

\hline
\textbf{Layer} & \textbf{Precision} & \textbf{Recall} & \textbf{F1} \\
\hline


Layer1 & 0.8136 (1090\%) &0.9013 (80\%) &0.8483 (605\%)\\
\hline

Layer2 & 0.8243 (49\%) &0.9150 (62\%) &0.8603 (185\%)\\
\hline

Layer3 & 0.8264 (33\%) &0.9173 (28\%) &0.8625 (38\%)\\
\hline

Layer4 & 0.8333 (34\%) &0.9170 (25\%) &0.8674 (40\%)\\
\hline

Layer5 & 0.8331 (32\%) &0.9211 (24\%) &0.8686 (37\%)\\
\hline

Layer6 & 0.8307 (30\%) &0.9211 (22\%) &0.8668 (23\%)\\
\hline

Layer7 & 0.8259 (29\%) &0.9201 (23\%) &0.8629 (33\%)\\
\hline

Layer8 & 0.8266 (31\%) &0.9198 (24\%) &0.8634 (35\%)\\
\hline

Layer9 & 0.8243 (31\%) &0.9182 (24\%) &0.8613 (35\%)\\
\hline

Layer10 & 0.8293 (30\%) &0.9182 (22\%) &0.8649 (34\%)\\
\hline

Layer11 & 0.8207 (15\%) &0.9201 (11\%) &0.8591 (15\%)\\
\hline

Layer12 & 0.8365 (11\%) &0.9240 (7\%) &0.8719 (10\%)\\
\hline

Layer-Pool & 0.8405 (4\%) &0.9260 (3\%) &0.8754 (4\%)\\
\hline

\end{tabular}
}
\end{table}

\textcolor{black}{To understand the semantic comprehension abilities of each layer, we extract the output hidden representation of the \texttt{[CLS]} token starting from Layer-1 until Layer-12. 
As explained previously, the embedding of \texttt{[CLS]} token captures the information about the input sequence and is used in classification tasks.
The linear head then uses the extracted hidden state over the BERT encoder.} The fine-tuning results showing the predictions of BERT for code clones are shown in Figure \ref{fig:clone}. 

We evaluate the performance of the code clone detection using evaluation metrics for classification tasks: Precision, Recall, and F1 score, as defined in previous sections. 

As seen in Figure \ref{fig:clone}, the lower layers have lower performance compared to the upper-level layers. This means that upper layers are better at representing the semantic relationship of the code. Therefore, semantic tasks should use upper-level layers to represent code to achieve the best performance. Interestingly, the performance remains uniform in the middle layers (4-10). This can be related to the uniform attention of identifiers, values we see in the Figure \ref{fig:attidf}. 

\begin{table}[htb]
\centering
\caption{Using CodeBERT \textit{identifier} embedding for code clone detection}
\label{table:robertaidf}
\small
\begin{tabular}{ccccl}

\hline
\textbf{Layer} & \textbf{Precision} & \textbf{Recall} & \textbf{F1} \\
\hline
CodeBERT\textsubscript{[CLS]} & 0.6337  &0.7632  &0.6294 \\
CodeBERT\textsubscript{IDF} & 0.7263 
&0.8631 
&0.7619 
\\

CodeBERT-MLM\textsubscript{[CLS]} & 0.5101  &0.8369  &0.6339 \\
CodeBERT-MLM\textsubscript{IDF} &0.6887 
& 0.9319 
& 0.7921 
\\
\hline
\end{tabular}
\end{table}

\subsection{Semantic Representation of Code for Code Clone Detection Using Identifiers} \label{IDF-experiment}

 Current works \cite{codebert, graphcodebert, ase-ptm} use the representation of the special tokens, i.e. \texttt{[CLS]} for code representation in downstream tasks for further fine-tuning. 
 However, based on the attention distribution we saw in sections \ref{rq1-results} and \ref{rq2-results}, we find that most of the BERT's attention is on the \textit{identifiers} in code, and the attention on \texttt{[CLS]} token is very low. In this part, we evaluate representing code using the identifiers' embeddings.
We perform an experiment similar to the one in Section \ref{CLS-CCD} on the performance of code clone detection for each layer, instead, using the \texttt{IDF} representation for code.
To incorporate the representation of identifiers, we calculate a weighted sum of all the identifiers present within the code. Each identifier embedding is normalized with its attention value and then added to form the weighted attention embedding of the identifiers. Finally, we extract the weighted embedding of the identifier from each layer and use it inside the classification head.
Table \ref{table:idf-clone} represents the results for each layer. 
The number inside parenthesis shows the performance change percentage compared to the model that uses representations of the \texttt{[CLS]} token.
The results show an increase in the performance for all metrics at each layer. 
Using the \texttt{IDF} representation increases the scores for the initial layers significantly, making the difference in the results of the lower and upper layers very small. 
Comparing this to what we saw in Figure \ref{fig:clone}, the Precision and Recall scores are below 75 for Layer-1 to Layer-11.  
The performance of the pooled layer is also increased by 3--4\% for all scores.

To evaluate the benefits of using embeddings from \texttt{IDF} over \texttt{[CLS]}, we repeat our experiments on CodeBERT \cite{codebert}. 
CodeBERT is chosen as it is a pre-trained model on code and it is previously evaluated on the same task by its authors, using the \texttt{[CLS]} token.
The experiments are conducted on both variations of CodeBERT, namely, CodeBERT-MLM and CodeBERT (which considers two training objectives) \cite{codebert}. We use the CodeBERT models available on HuggingFace and use the same code made available by the authors for fine-tuning on code clone detection. 
Similar to the pre-processing we applied for BERT to pre-trained it on code, we tokenize the code first using the JavaLang tokenizer. 
The tokenized code is used for code clone detection. This tokenization is necessary to extract and map the tokens to their respective syntactic types. 
We experimented only for the pooled layer of CodeBERT as it was the best performing model in our experiments conducted on BERT.
The results are shown in Table \ref{table:robertaidf}, where the subscript to the models' name indicates the representation of the token that is used. 
When embeddings from identifiers are used, the performance of CodeBERT is increased for both variations of CodeBERT. 
CodeBERT results are improved by 14.61\%, 13.08\%, and 21.05\% for Precision, Recall, and F1 scores, respectively. The three metrics in order are 35.01\%, 11.035\%, and 24.95\% for CodeBERT-MLM. 
Note that CodeBERT is based on RoBERTa \cite{roberta}, which is a Transformer model based on BERT.

{\textbf{RQ3 Summary:} BERT can learn the semantic characteristics of code. The lower layers contain semantic knowledge, whereas the upper layers in BERT are rich in semantic context. Employing embeddings generated from \textit{identifiers} in place of \texttt{[CLS]} token for the code clone detection task can boost the performance of the model by 4\% in the pooled layer and achieve results in the initial layers (0.848 F1 score), which are close to the final results (0.875 F1 score). This finding is not bound to BERT and improves the performance of CodeBERT (by 21--24\% F1 score), which is based on RoBERTa when using \texttt{IDF} embeddings over \texttt{[CLS]}. }

\section{Related Work} \label{related-work}
Attention-based neural networks are extensively used in software engineering \cite{attend, code2seq}. Attention networks within the deep learning models help the model focus more on the important tokens, which can eventually affect the predictions made by the model. An attention mechanism would assign higher attention values/energy to important entities inside the input and its analysis or value in different studies to build a model or explain its predictions. 
LeClair et al. \cite{attend} provide attention heat maps for the predictions of their model for code summarization to explain the prediction based on the amount of attention the model puts on relevant tokens. 
Alon et al. \cite{code2seq} use an attention framework to select import Abstract Syntax Tree paths between the node to predict the appropriate function names.
Similarly, Haque et al. \cite{msr-context} leverage the attention framework to select relevant parts of the context file to generate summaries. 
Shuai et al. \cite{shuai} learn interdependent representations for code and query with an attention mechanism for code search, and Li et al. \cite{li} leverage attention-based network for bug detection.
Unlike previous works, we do not leverage the attention mechanism to train a neural model; rather explore the attention mechanism and analyze it for code. 

Literature for analyzing the behavior of PLMs is limited. Karmakar et al. \cite{new-ptm-study} studies the comprehension abilities of four different pre-trained models trained on code namely, BERT\cite{bert}, CodeBERTa\cite{codeberta}, CodeBERT\cite{codebert} and GraphCodeBERT. They compare the performance of these models from the perspective of probing classifiers. Other studies \cite{new-1, new-2, new-3, new-4}, show that identifiers are important code entities and can be used in the modeling of Transfoemr based models.  This work presents the first study in software engineering, which analyses the multi-headed attention framework of BERT, which is not done previously \cite{new-ptm-study}. Another difference between this work and \cite{new-ptm-study} is the use of different probing classifiers. We use code clone detection and non-parametric probing for syntax tagging of code whereas in \cite{new-ptm-study}, it uses other probing tasks. 
Also, they compare the semantic and syntactic \textit{performance of different PLMs}, but we explore the \textit{semantic and syntax abilities of BERT with respect to what it learns about code constructs}. 

\textbf{Differences among our work and previous studies:}
Our study analyses where and how attention exists within the pre-trained models for code and what kind of information, both semantic and syntactic, is learned by the model. The current research recognizes the success and importance of pre-trained models; however, the software engineering research community has not focused on understanding the comprehension abilities of the PLMs for programming languages. 
\textcolor{black}{The current studies do not generate any new embedding using identifiers, do not draw comparison with NLP, and do not study the attention mechanism of the models.}
A body of research in NLP focus on exploring the internal working of the PLMs on English. Our work is inspired and is closer to \cite{whatdoes, nlppipeline2, distributional, distributional2}. The difference between our work and these studies is on different domains: software engineering and NLP. We explore BERT's semantic and syntactic behavior with respect to code constructs and assess our findings on code clone detection. 

\section{Threats to Validity} \label{threat}
\textbf{Internal threats} in this study can be related to inadequate training of the models. To eliminate this threat, we used the official implementation of BERT that is made available by its authors. The results obtained for training BERT on our dataset are similar to the original results of the BERT model reported by the authors of BERT, thus, ensuring the proper training of the model. Moreover, for CodeBERT, we use its official pre-trained weights provided by the authors, mitigating threats related to the training of CodeBERT. 

\textbf{External Threats} relate to the dataset used in our work. We use a popular open-source dataset, CodeSearchNet \cite{codesearchnet}.
The same dataset has been used in multiple PLM studies for code \cite{codebert, graphcodebert, codexglue}. As the dataset is commonly used for pre-training PLMs, we anticipate low threats related to the choice of the dataset. Another threat is related to the chosen task. We only applied our analysis on code clone detection. Although more studies are needed, our results shed lights for other tasks as code clone detection uses both semantic and syntactic knowledge about the code. We only study Java, and the results might not be generalizable to other languages.

\textbf{Conclusion Threats} in our work refer to the threats caused by bias used in sampling code from the dataset. To mitigate this threat, we apply random sampling, and the experiments are repeated for both cased and uncased code sequences. The same results were obtained in both cases. Another bias could be related to the choice of model. To reduce this threat, we applied the identifier embedding for code clone detection on two different models, BERT and CodeBERT.. 

\section{Conclusion and Future Work} \label{conclusion}
In this work, we studied the attention behavior of BERT for code and compared it to the attention behavior in NLP studies.
We explored the relationship among the code constructs and found that identifiers have higher attention. An identifier attends towards itself more than other tokens; some of its attention value is made to its immediate left and right token rather than on the tokens farther from it. Based on this finding, we proposed to use the representation of identifiers in code clone detection, which requires the code semantic information. We showed this improves the clone detection results in all layers, and this finding is not restricted to BERT. 
The findings can be used to build smaller models which can then be used to employ the identifiers' embeddings in the downstream task (e.g. by models with 6 layers instead of 12 layers), to attain the similar performance to that of larger models. A future direction could be evaluating the results on other software engineering tasks requiring code intelligence. It would also be interesting to confirm whether the combination of the identifiers' and separators' embeddings would generate meaningful embeddings or introduce noise for code.

\begin{acks}
This research is support by a grant from Natural Sciences and Engineering Research Council of Canada RGPIN-2019-05175.
\end{acks}

\balance
\bibliographystyle{ACM-Reference-Format}
\bibliography{mainbib}


\begin{thebibliography}{50}


\ifx \showCODEN    \undefined \def \showCODEN     #1{\unskip}     \fi
\ifx \showDOI      \undefined \def \showDOI       #1{#1}\fi
\ifx \showISBNx    \undefined \def \showISBNx     #1{\unskip}     \fi
\ifx \showISBNxiii \undefined \def \showISBNxiii  #1{\unskip}     \fi
\ifx \showISSN     \undefined \def \showISSN      #1{\unskip}     \fi
\ifx \showLCCN     \undefined \def \showLCCN      #1{\unskip}     \fi
\ifx \shownote     \undefined \def \shownote      #1{#1}          \fi
\ifx \showarticletitle \undefined \def \showarticletitle #1{#1}   \fi
\ifx \showURL      \undefined \def \showURL       {\relax}        \fi
\providecommand\bibfield[2]{#2}
\providecommand\bibinfo[2]{#2}
\providecommand\natexlab[1]{#1}
\providecommand\showeprint[2][]{arXiv:#2}

\bibitem[\protect\citeauthoryear{Ahmad, Chakraborty, Ray, and Chang}{Ahmad
  et~al\mbox{.}}{2021}]%
        {plbart}
\bibfield{author}{\bibinfo{person}{Wasi~Uddin Ahmad}, \bibinfo{person}{Saikat
  Chakraborty}, \bibinfo{person}{Baishakhi Ray}, {and}
  \bibinfo{person}{Kai{-}Wei Chang}.} \bibinfo{year}{2021}\natexlab{}.
\newblock In \bibinfo{booktitle}{\emph{Proceedings of the 2021 Conference of
  the North American Chapter of the Association for Computational Linguistics:
  Human Language Technologies}}.
\newblock


\bibitem[\protect\citeauthoryear{Ahmad, Chakraborty, Ray, and Chang}{Ahmad
  et~al\mbox{.}}{2020}]%
        {transformer}
\bibfield{author}{\bibinfo{person}{Wasi~Uddin Ahmad}, \bibinfo{person}{Saikat
  Chakraborty}, \bibinfo{person}{Baishakhi Ray}, {and} \bibinfo{person}{Kai-Wei
  Chang}.} \bibinfo{year}{2020}\natexlab{}.
\newblock \showarticletitle{A Transformer-based Approach for Source Code
  Summarization}. In \bibinfo{booktitle}{\emph{ACL}}.
  \bibinfo{pages}{4998--5007}.
\newblock
\urldef\tempurl%
\url{https://www.aclweb.org/anthology/2020.acl-main.449/}
\showURL{%
\tempurl}


\bibitem[\protect\citeauthoryear{Alon, Brody, Levy, and Yahav}{Alon
  et~al\mbox{.}}{2019}]%
        {code2seq}
\bibfield{author}{\bibinfo{person}{Uri Alon}, \bibinfo{person}{Shaked Brody},
  \bibinfo{person}{Omer Levy}, {and} \bibinfo{person}{Eran Yahav}.}
  \bibinfo{year}{2019}\natexlab{}.
\newblock \showarticletitle{code2seq: Generating Sequences from Structured
  Representations of Code}. In \bibinfo{booktitle}{\emph{7th International
  Conference on Learning Representations, {ICLR} 2019, New Orleans, LA, USA,
  May 6-9, 2019}}.
\newblock


\bibitem[\protect\citeauthoryear{Alsuhaibani, Newman, Decker, Collard, and
  Maletic}{Alsuhaibani et~al\mbox{.}}{2021}]%
        {idf3}
\bibfield{author}{\bibinfo{person}{Reem~S. Alsuhaibani},
  \bibinfo{person}{Christian~D. Newman}, \bibinfo{person}{Michael~J. Decker},
  \bibinfo{person}{Michael~L. Collard}, {and} \bibinfo{person}{Jonathan~I.
  Maletic}.} \bibinfo{year}{2021}\natexlab{}.
\newblock \showarticletitle{On the Naming of Methods: A Survey of Professional
  Developers}. In \bibinfo{booktitle}{\emph{2021 IEEE/ACM 43rd International
  Conference on Software Engineering (ICSE)}}. \bibinfo{pages}{587--599}.
\newblock
\urldef\tempurl%
\url{https://doi.org/10.1109/ICSE43902.2021.00061}
\showDOI{\tempurl}


\bibitem[\protect\citeauthoryear{Applis, Panichella, and van Deursen}{Applis
  et~al\mbox{.}}{2021}]%
        {new-3}
\bibfield{author}{\bibinfo{person}{Leonhard Applis}, \bibinfo{person}{Annibale
  Panichella}, {and} \bibinfo{person}{Arie van Deursen}.}
  \bibinfo{year}{2021}\natexlab{}.
\newblock \showarticletitle{Assessing Robustness of ML-Based Program Analysis
  Tools using Metamorphic Program Transformations}. In
  \bibinfo{booktitle}{\emph{2021 36th IEEE/ACM International Conference on
  Automated Software Engineering (ASE)}}. IEEE, \bibinfo{pages}{1377--1381}.
\newblock


\bibitem[\protect\citeauthoryear{Baxter, Yahin, Moura, Sant'Anna, and
  Bier}{Baxter et~al\mbox{.}}{1998}]%
        {clone1}
\bibfield{author}{\bibinfo{person}{I.D. Baxter}, \bibinfo{person}{A. Yahin},
  \bibinfo{person}{L. Moura}, \bibinfo{person}{M. Sant'Anna}, {and}
  \bibinfo{person}{L. Bier}.} \bibinfo{year}{1998}\natexlab{}.
\newblock \showarticletitle{Clone detection using abstract syntax trees}. In
  \bibinfo{booktitle}{\emph{Proceedings. International Conference on Software
  Maintenance (Cat. No. 98CB36272)}}. \bibinfo{pages}{368--377}.
\newblock
\urldef\tempurl%
\url{https://doi.org/10.1109/ICSM.1998.738528}
\showDOI{\tempurl}


\bibitem[\protect\citeauthoryear{Casalnuovo, Sagae, and Devanbu}{Casalnuovo
  et~al\mbox{.}}{2019}]%
        {difnlpl}
\bibfield{author}{\bibinfo{person}{Casey Casalnuovo}, \bibinfo{person}{Kenji
  Sagae}, {and} \bibinfo{person}{Premkumar~T. Devanbu}.}
  \bibinfo{year}{2019}\natexlab{}.
\newblock \showarticletitle{Studying the Difference Between Natural and
  Programming Language Corpora}.
\newblock \bibinfo{journal}{\emph{Empirical Software Engineering}}
  (\bibinfo{year}{2019}).
\newblock


\bibitem[\protect\citeauthoryear{Clark, Khandelwal, Levy, and Manning}{Clark
  et~al\mbox{.}}{2019}]%
        {whatdoes}
\bibfield{author}{\bibinfo{person}{Kevin Clark}, \bibinfo{person}{Urvashi
  Khandelwal}, \bibinfo{person}{Omer Levy}, {and}
  \bibinfo{person}{Christopher~D. Manning}.} \bibinfo{year}{2019}\natexlab{}.
\newblock \showarticletitle{What Does {BERT} Look At? An Analysis of BERT's
  Attention}. In \bibinfo{booktitle}{\emph{Proceedings of the 2019 ACL Workshop
  BlackboxNLP: Analyzing and Interpreting Neural Networks for NLP}}.
\newblock


\bibitem[\protect\citeauthoryear{Compton, Frank, Patros, and Koay}{Compton
  et~al\mbox{.}}{2020}]%
        {new-1}
\bibfield{author}{\bibinfo{person}{Rhys Compton}, \bibinfo{person}{Eibe Frank},
  \bibinfo{person}{Panos Patros}, {and} \bibinfo{person}{Abigail Koay}.}
  \bibinfo{year}{2020}\natexlab{}.
\newblock \showarticletitle{Embedding Java Classes with code2vec}.
\newblock \bibinfo{journal}{\emph{Proceedings of the 17th International
  Conference on Mining Software Repositories}} (\bibinfo{date}{Jun}
  \bibinfo{year}{2020}).
\newblock
\urldef\tempurl%
\url{https://doi.org/10.1145/3379597.3387445}
\showDOI{\tempurl}


\bibitem[\protect\citeauthoryear{Devlin, Chang, Lee, and Toutanova}{Devlin
  et~al\mbox{.}}{2019}]%
        {bert}
\bibfield{author}{\bibinfo{person}{Jacob Devlin}, \bibinfo{person}{Ming-Wei
  Chang}, \bibinfo{person}{Kenton Lee}, {and} \bibinfo{person}{Kristina
  Toutanova}.} \bibinfo{year}{2019}\natexlab{}.
\newblock \showarticletitle{{BERT}: Pre-training of Deep Bidirectional
  Transformers for Language Understanding}. In
  \bibinfo{booktitle}{\emph{Proceedings of the 2019 Conference of the North
  {A}merican Chapter of the Association for Computational Linguistics: Human
  Language Technologies, Volume 1 (Long and Short Papers)}}.
  \bibinfo{publisher}{Association for Computational Linguistics},
  \bibinfo{address}{Minneapolis, Minnesota}, \bibinfo{pages}{4171--4186}.
\newblock
\urldef\tempurl%
\url{https://doi.org/10.18653/v1/N19-1423}
\showDOI{\tempurl}


\bibitem[\protect\citeauthoryear{Feitelson, Mizrahi, Noy, Ben~Shabat, Eliyahu,
  and Sheffer}{Feitelson et~al\mbox{.}}{2020}]%
        {idf2}
\bibfield{author}{\bibinfo{person}{Dror Feitelson}, \bibinfo{person}{Ayelet
  Mizrahi}, \bibinfo{person}{Nofar Noy}, \bibinfo{person}{Aviad Ben~Shabat},
  \bibinfo{person}{Or Eliyahu}, {and} \bibinfo{person}{Roy Sheffer}.}
  \bibinfo{year}{2020}\natexlab{}.
\newblock \showarticletitle{How Developers Choose Names}.
\newblock \bibinfo{journal}{\emph{IEEE Transactions on Software Engineering}}
  (\bibinfo{year}{2020}), \bibinfo{pages}{1--1}.
\newblock
\urldef\tempurl%
\url{https://doi.org/10.1109/TSE.2020.2976920}
\showDOI{\tempurl}


\bibitem[\protect\citeauthoryear{Feng, Guo, Tang, Duan, Feng, Gong, Shou, Qin,
  Liu, Jiang, and Zhou}{Feng et~al\mbox{.}}{2020}]%
        {codebert}
\bibfield{author}{\bibinfo{person}{Zhangyin Feng}, \bibinfo{person}{Daya Guo},
  \bibinfo{person}{Duyu Tang}, \bibinfo{person}{Nan Duan},
  \bibinfo{person}{Xiaocheng Feng}, \bibinfo{person}{Ming Gong},
  \bibinfo{person}{Linjun Shou}, \bibinfo{person}{Bing Qin},
  \bibinfo{person}{Ting Liu}, \bibinfo{person}{Daxin Jiang}, {and}
  \bibinfo{person}{Ming Zhou}.} \bibinfo{year}{2020}\natexlab{}.
\newblock \showarticletitle{{C}ode{BERT}: A Pre-Trained Model for Programming
  and Natural Languages}. In \bibinfo{booktitle}{\emph{Findings of the
  Association for Computational Linguistics: EMNLP 2020}}.
  \bibinfo{publisher}{Association for Computational Linguistics},
  \bibinfo{address}{Online}, \bibinfo{pages}{1536--1547}.
\newblock
\urldef\tempurl%
\url{https://doi.org/10.18653/v1/2020.findings-emnlp.139}
\showDOI{\tempurl}


\bibitem[\protect\citeauthoryear{Gao, Gao, He, Zeng, Nie, and Xia}{Gao
  et~al\mbox{.}}{2021}]%
        {xia-trans}
\bibfield{author}{\bibinfo{person}{Shuzheng Gao}, \bibinfo{person}{Cuiyun Gao},
  \bibinfo{person}{Yulan He}, \bibinfo{person}{Jichuan Zeng},
  \bibinfo{person}{Lun~Yiu Nie}, {and} \bibinfo{person}{Xin Xia}.}
  \bibinfo{year}{2021}\natexlab{}.
\newblock \showarticletitle{Code Structure Guided Transformer for Source Code
  Summarization}.
\newblock \bibinfo{journal}{\emph{CoRR}}  \bibinfo{volume}{abs/2104.09340}
  (\bibinfo{year}{2021}).
\newblock
\showeprint[arxiv]{2104.09340}
\urldef\tempurl%
\url{https://arxiv.org/abs/2104.09340}
\showURL{%
\tempurl}


\bibitem[\protect\citeauthoryear{Guo, Ren, Lu, Feng, Tang, Liu, Zhou, Duan,
  Svyatkovskiy, Fu, Tufano, Deng, Clement, Drain, Sundaresan, Yin, Jiang, and
  Zhou}{Guo et~al\mbox{.}}{2021}]%
        {graphcodebert}
\bibfield{author}{\bibinfo{person}{Daya Guo}, \bibinfo{person}{Shuo Ren},
  \bibinfo{person}{Shuai Lu}, \bibinfo{person}{Zhangyin Feng},
  \bibinfo{person}{Duyu Tang}, \bibinfo{person}{Shujie Liu},
  \bibinfo{person}{Long Zhou}, \bibinfo{person}{Nan Duan},
  \bibinfo{person}{Alexey Svyatkovskiy}, \bibinfo{person}{Shengyu Fu},
  \bibinfo{person}{Michele Tufano}, \bibinfo{person}{Shao~Kun Deng},
  \bibinfo{person}{Colin Clement}, \bibinfo{person}{Dawn Drain},
  \bibinfo{person}{Neel Sundaresan}, \bibinfo{person}{Jian Yin},
  \bibinfo{person}{Daxin Jiang}, {and} \bibinfo{person}{Ming Zhou}.}
  \bibinfo{year}{2021}\natexlab{}.
\newblock \showarticletitle{GraphCodeBERT: Pre-training Code Representations
  with Data Flow}. In \bibinfo{booktitle}{\emph{International Conference on
  Learning Representations}}.
\newblock


\bibitem[\protect\citeauthoryear{Haque, LeClair, Wu, and McMillan}{Haque
  et~al\mbox{.}}{2020}]%
        {msr-context}
\bibfield{author}{\bibinfo{person}{Sakib Haque}, \bibinfo{person}{Alexander
  LeClair}, \bibinfo{person}{Lingfei Wu}, {and} \bibinfo{person}{Collin
  McMillan}.} \bibinfo{year}{2020}\natexlab{}.
\newblock \showarticletitle{Improved Automatic Summarization of Subroutines via
  Attention to File Context}. In \bibinfo{booktitle}{\emph{Proceedings of the
  17th International Conference on Mining Software Repositories}} (Seoul,
  Republic of Korea) \emph{(\bibinfo{series}{MSR '20})}.
  \bibinfo{publisher}{Association for Computing Machinery},
  \bibinfo{address}{New York, NY, USA}, \bibinfo{pages}{300–310}.
\newblock
\showISBNx{9781450375177}
\urldef\tempurl%
\url{https://doi.org/10.1145/3379597.3387449}
\showDOI{\tempurl}


\bibitem[\protect\citeauthoryear{Harer, Kim, Russell, Ozdemir, Kosta,
  Rangamani, Hamilton, Centeno, Key, Ellingwood, Antelman, Mackay, McConley,
  Opper, Chin, and Lazovich}{Harer et~al\mbox{.}}{2018}]%
        {cv2}
\bibfield{author}{\bibinfo{person}{Jacob~A. Harer}, \bibinfo{person}{Louis~Y.
  Kim}, \bibinfo{person}{Rebecca~L. Russell}, \bibinfo{person}{Onur Ozdemir},
  \bibinfo{person}{Leonard~R. Kosta}, \bibinfo{person}{Akshay Rangamani},
  \bibinfo{person}{Lei~H. Hamilton}, \bibinfo{person}{Gabriel~I. Centeno},
  \bibinfo{person}{Jonathan~R. Key}, \bibinfo{person}{Paul~M. Ellingwood},
  \bibinfo{person}{Erik Antelman}, \bibinfo{person}{Alan Mackay},
  \bibinfo{person}{Marc~W. McConley}, \bibinfo{person}{Jeffrey~M. Opper},
  \bibinfo{person}{Peter Chin}, {and} \bibinfo{person}{Tomo Lazovich}.}
  \bibinfo{year}{2018}\natexlab{}.
\newblock \showarticletitle{Automated software vulnerability detection with
  machine learning}.
\newblock \bibinfo{journal}{\emph{arXiv e-prints}}, Article
  \bibinfo{articleno}{arXiv:1803.04497} (\bibinfo{date}{feb}
  \bibinfo{year}{2018}), \bibinfo{numpages}{arXiv:1803.04497}~pages.
\newblock
\showeprint[arxiv]{1803.04497}~[cs.SE]


\bibitem[\protect\citeauthoryear{Hu, Li, Xia, Lo, and Jin}{Hu
  et~al\mbox{.}}{2018}]%
        {hu1}
\bibfield{author}{\bibinfo{person}{Xing Hu}, \bibinfo{person}{Ge Li},
  \bibinfo{person}{Xin Xia}, \bibinfo{person}{David Lo}, {and}
  \bibinfo{person}{Zhi Jin}.} \bibinfo{year}{2018}\natexlab{}.
\newblock \showarticletitle{Deep Code Comment Generation}. In
  \bibinfo{booktitle}{\emph{Proceedings of the 26th Conference on Program
  Comprehension}} (Gothenburg, Sweden) \emph{(\bibinfo{series}{ICPC '18})}.
  \bibinfo{publisher}{Association for Computing Machinery},
  \bibinfo{address}{New York, NY, USA}, \bibinfo{pages}{200–210}.
\newblock
\showISBNx{9781450357142}
\urldef\tempurl%
\url{https://doi.org/10.1145/3196321.3196334}
\showDOI{\tempurl}


\bibitem[\protect\citeauthoryear{Husain, Wu, Gazit, Allamanis, and
  Brockschmidt}{Husain et~al\mbox{.}}{2019}]%
        {codesearchnet}
\bibfield{author}{\bibinfo{person}{Hamel Husain}, \bibinfo{person}{Ho-Hsiang
  Wu}, \bibinfo{person}{Tiferet Gazit}, \bibinfo{person}{Miltiadis Allamanis},
  {and} \bibinfo{person}{Marc Brockschmidt}.} \bibinfo{year}{2019}\natexlab{}.
\newblock \showarticletitle{{CodeSearchNet Challenge: Evaluating the State of
  Semantic Code Search}}.
\newblock \bibinfo{journal}{\emph{arXiv e-prints}}, Article
  \bibinfo{articleno}{arXiv:1909.09436} (\bibinfo{date}{Sept.}
  \bibinfo{year}{2019}), \bibinfo{numpages}{arXiv:1909.09436}~pages.
\newblock
\showeprint[arxiv]{1909.09436}~[cs.LG]


\bibitem[\protect\citeauthoryear{Kanade, Maniatis, Balakrishnan, and
  Shi}{Kanade et~al\mbox{.}}{2020}]%
        {cubert}
\bibfield{author}{\bibinfo{person}{Aditya Kanade}, \bibinfo{person}{Petros
  Maniatis}, \bibinfo{person}{Gogul Balakrishnan}, {and}
  \bibinfo{person}{Kensen Shi}.} \bibinfo{year}{2020}\natexlab{}.
\newblock \showarticletitle{Learning and Evaluating Contextual Embedding of
  Source Code}. In \bibinfo{booktitle}{\emph{Proceedings of the 37th
  International Conference on Machine Learning}}
  \emph{(\bibinfo{series}{Proceedings of Machine Learning Research},
  Vol.~\bibinfo{volume}{119})}, \bibfield{editor}{\bibinfo{person}{Hal~Daumé
  III} {and} \bibinfo{person}{Aarti Singh}} (Eds.). \bibinfo{publisher}{PMLR},
  \bibinfo{address}{Virtual}, \bibinfo{pages}{5110--5121}.
\newblock
\urldef\tempurl%
\url{http://proceedings.mlr.press/v119/kanade20a.html}
\showURL{%
\tempurl}


\bibitem[\protect\citeauthoryear{Karampatsis and Sutton}{Karampatsis and
  Sutton}{2020}]%
        {scelmo}
\bibfield{author}{\bibinfo{person}{Rafael-Michael Karampatsis} {and}
  \bibinfo{person}{Charles Sutton}.} \bibinfo{year}{2020}\natexlab{}.
\newblock \showarticletitle{{SCELMo: Source Code Embeddings from Language
  Models}}.
\newblock \bibinfo{journal}{\emph{arXiv e-prints}}, Article
  \bibinfo{articleno}{arXiv:2004.13214} (\bibinfo{date}{April}
  \bibinfo{year}{2020}), \bibinfo{numpages}{arXiv:2004.13214}~pages.
\newblock
\showeprint[arxiv]{2004.13214}~[cs.SE]


\bibitem[\protect\citeauthoryear{Karmakar and Robbes}{Karmakar and
  Robbes}{2021}]%
        {new-ptm-study}
\bibfield{author}{\bibinfo{person}{Anjan Karmakar} {and}
  \bibinfo{person}{Romain Robbes}.} \bibinfo{year}{2021}\natexlab{}.
\newblock \showarticletitle{What do pre-trained code models know about code?}
\newblock \bibinfo{journal}{\emph{CoRR}}  \bibinfo{volume}{abs/2108.11308}
  (\bibinfo{year}{2021}).
\newblock
\showeprint[arXiv]{2108.11308}
\urldef\tempurl%
\url{https://arxiv.org/abs/2108.11308}
\showURL{%
\tempurl}


\bibitem[\protect\citeauthoryear{Krinke}{Krinke}{2001}]%
        {clone2}
\bibfield{author}{\bibinfo{person}{J. Krinke}.}
  \bibinfo{year}{2001}\natexlab{}.
\newblock \showarticletitle{Identifying similar code with program dependence
  graphs}. In \bibinfo{booktitle}{\emph{Proceedings Eighth Working Conference
  on Reverse Engineering}}. \bibinfo{pages}{301--309}.
\newblock
\urldef\tempurl%
\url{https://doi.org/10.1109/WCRE.2001.957835}
\showDOI{\tempurl}


\bibitem[\protect\citeauthoryear{LeClair, Jiang, and McMillan}{LeClair
  et~al\mbox{.}}{2019a}]%
        {attend}
\bibfield{author}{\bibinfo{person}{Alexander LeClair}, \bibinfo{person}{Siyuan
  Jiang}, {and} \bibinfo{person}{Collin McMillan}.}
  \bibinfo{year}{2019}\natexlab{a}.
\newblock \showarticletitle{A Neural Model for Generating Natural Language
  Summaries of Program Subroutines}. In \bibinfo{booktitle}{\emph{2019 IEEE/ACM
  41st International Conference on Software Engineering (ICSE)}}.
\newblock


\bibitem[\protect\citeauthoryear{LeClair, Jiang, and McMillan}{LeClair
  et~al\mbox{.}}{2019b}]%
        {clair}
\bibfield{author}{\bibinfo{person}{Alexander LeClair}, \bibinfo{person}{Siyuan
  Jiang}, {and} \bibinfo{person}{Collin McMillan}.}
  \bibinfo{year}{2019}\natexlab{b}.
\newblock \showarticletitle{A Neural Model for Generating Natural Language
  Summaries of Program Subroutines}. In \bibinfo{booktitle}{\emph{2019 IEEE/ACM
  41st International Conference on Software Engineering (ICSE)}}.
\newblock


\bibitem[\protect\citeauthoryear{Li, Wang, Nguyen, and Van~Nguyen}{Li
  et~al\mbox{.}}{2019}]%
        {li}
\bibfield{author}{\bibinfo{person}{Yi Li}, \bibinfo{person}{Shaohua Wang},
  \bibinfo{person}{Tien~N. Nguyen}, {and} \bibinfo{person}{Son Van~Nguyen}.}
  \bibinfo{year}{2019}\natexlab{}.
\newblock \showarticletitle{Improving Bug Detection via Context-Based Code
  Representation Learning and Attention-Based Neural Networks}.
\newblock \bibinfo{journal}{\emph{Proc. ACM Program. Lang.}}
  \bibinfo{volume}{3}, \bibinfo{number}{OOPSLA}, Article
  \bibinfo{articleno}{162} (\bibinfo{date}{oct} \bibinfo{year}{2019}),
  \bibinfo{numpages}{30}~pages.
\newblock
\urldef\tempurl%
\url{https://doi.org/10.1145/3360588}
\showDOI{\tempurl}


\bibitem[\protect\citeauthoryear{Liu, Li, Zhao, and Jin}{Liu
  et~al\mbox{.}}{2020}]%
        {ase-ptm}
\bibfield{author}{\bibinfo{person}{Fang Liu}, \bibinfo{person}{Ge Li},
  \bibinfo{person}{Yunfei Zhao}, {and} \bibinfo{person}{Zhi Jin}.}
  \bibinfo{year}{2020}\natexlab{}.
\newblock \showarticletitle{Multi-task Learning based Pre-trained Language
  Model for Code Completion}. In \bibinfo{booktitle}{\emph{2020 35th IEEE/ACM
  International Conference on Automated Software Engineering (ASE)}}.
  \bibinfo{pages}{473--485}.
\newblock


\bibitem[\protect\citeauthoryear{Liu, Ott, Goyal, Du, Joshi, Chen, Levy, Lewis,
  Zettlemoyer, and Stoyanov}{Liu et~al\mbox{.}}{2019}]%
        {roberta}
\bibfield{author}{\bibinfo{person}{Yinhan Liu}, \bibinfo{person}{Myle Ott},
  \bibinfo{person}{Naman Goyal}, \bibinfo{person}{Jingfei Du},
  \bibinfo{person}{Mandar Joshi}, \bibinfo{person}{Danqi Chen},
  \bibinfo{person}{Omer Levy}, \bibinfo{person}{Mike Lewis},
  \bibinfo{person}{Luke Zettlemoyer}, {and} \bibinfo{person}{Veselin
  Stoyanov}.} \bibinfo{year}{2019}\natexlab{}.
\newblock \showarticletitle{RoBERTa: {A} Robustly Optimized {BERT} Pretraining
  Approach}.
\newblock \bibinfo{journal}{\emph{CoRR}}  \bibinfo{volume}{abs/1907.11692}
  (\bibinfo{year}{2019}).
\newblock
\showeprint[arXiv]{1907.11692}
\urldef\tempurl%
\url{http://arxiv.org/abs/1907.11692}
\showURL{%
\tempurl}


\bibitem[\protect\citeauthoryear{Lu, Guo, Ren, Huang, Svyatkovskiy, Blanco,
  Clement, Drain, Jiang, Tang, Li, Zhou, Shou, Zhou, Tufano, Gong, Zhou, Duan,
  Sundaresan, Deng, Fu, and Liu}{Lu et~al\mbox{.}}{2021}]%
        {codexglue}
\bibfield{author}{\bibinfo{person}{Shuai Lu}, \bibinfo{person}{Daya Guo},
  \bibinfo{person}{Shuo Ren}, \bibinfo{person}{Junjie Huang},
  \bibinfo{person}{Alexey Svyatkovskiy}, \bibinfo{person}{Ambrosio Blanco},
  \bibinfo{person}{Colin~B. Clement}, \bibinfo{person}{Dawn Drain},
  \bibinfo{person}{Daxin Jiang}, \bibinfo{person}{Duyu Tang},
  \bibinfo{person}{Ge Li}, \bibinfo{person}{Lidong Zhou},
  \bibinfo{person}{Linjun Shou}, \bibinfo{person}{Long Zhou},
  \bibinfo{person}{Michele Tufano}, \bibinfo{person}{Ming Gong},
  \bibinfo{person}{Ming Zhou}, \bibinfo{person}{Nan Duan},
  \bibinfo{person}{Neel Sundaresan}, \bibinfo{person}{Shao~Kun Deng},
  \bibinfo{person}{Shengyu Fu}, {and} \bibinfo{person}{Shujie Liu}.}
  \bibinfo{year}{2021}\natexlab{}.
\newblock \showarticletitle{CodeXGLUE: {A} Machine Learning Benchmark Dataset
  for Code Understanding and Generation}.
\newblock \bibinfo{journal}{\emph{CoRR}}  \bibinfo{volume}{abs/2102.04664}
  (\bibinfo{year}{2021}).
\newblock
\showeprint[arxiv]{2102.04664}
\urldef\tempurl%
\url{https://arxiv.org/abs/2102.04664}
\showURL{%
\tempurl}


\bibitem[\protect\citeauthoryear{Malik, Patra, and Pradel}{Malik
  et~al\mbox{.}}{2019}]%
        {type-pred}
\bibfield{author}{\bibinfo{person}{Rabee~S. Malik}, \bibinfo{person}{Jibesh
  Patra}, {and} \bibinfo{person}{Michael Pradel}.}
  \bibinfo{year}{2019}\natexlab{}.
\newblock \showarticletitle{NL2Type: Inferring JavaScript Function Types from
  Natural Language Information}. In \bibinfo{booktitle}{\emph{2019 IEEE/ACM
  41st International Conference on Software Engineering (ICSE)}}.
  \bibinfo{pages}{304--315}.
\newblock
\urldef\tempurl%
\url{https://doi.org/10.1109/ICSE.2019.00045}
\showDOI{\tempurl}


\bibitem[\protect\citeauthoryear{Manning, Clark, Hewitt, Khandelwal, and
  Levy}{Manning et~al\mbox{.}}{2020}]%
        {nlppipeline1}
\bibfield{author}{\bibinfo{person}{Christopher~D Manning},
  \bibinfo{person}{Kevin Clark}, \bibinfo{person}{John Hewitt},
  \bibinfo{person}{Urvashi Khandelwal}, {and} \bibinfo{person}{Omer Levy}.}
  \bibinfo{year}{2020}\natexlab{}.
\newblock \showarticletitle{Emergent linguistic structure in artificial neural
  networks trained by self-supervision}.
\newblock \bibinfo{journal}{\emph{Proceedings of the National Academy of
  Sciences}} \bibinfo{volume}{117}, \bibinfo{number}{48}
  (\bibinfo{year}{2020}), \bibinfo{pages}{30046--30054}.
\newblock


\bibitem[\protect\citeauthoryear{Menéndez, Pardo, Pardo, and Pardo}{Menéndez
  et~al\mbox{.}}{1997}]%
        {jsd}
\bibfield{author}{\bibinfo{person}{M.L. Menéndez}, \bibinfo{person}{J.A.
  Pardo}, \bibinfo{person}{L. Pardo}, {and} \bibinfo{person}{M.C. Pardo}.}
  \bibinfo{year}{1997}\natexlab{}.
\newblock \showarticletitle{The Jensen-Shannon divergence}.
\newblock \bibinfo{journal}{\emph{Journal of the Franklin Institute}}
  \bibinfo{volume}{334}, \bibinfo{number}{2} (\bibinfo{year}{1997}),
  \bibinfo{pages}{307--318}.
\newblock
\showISSN{0016-0032}
\urldef\tempurl%
\url{https://doi.org/10.1016/S0016-0032(96)00063-4}
\showDOI{\tempurl}


\bibitem[\protect\citeauthoryear{Newman, AlSuhaibani, Decker, Peruma, Kaushik,
  Mkaouer, and Hill}{Newman et~al\mbox{.}}{2020}]%
        {pos}
\bibfield{author}{\bibinfo{person}{Christian~D. Newman},
  \bibinfo{person}{Reem~S. AlSuhaibani}, \bibinfo{person}{Michael~J. Decker},
  \bibinfo{person}{Anthony Peruma}, \bibinfo{person}{Dishant Kaushik},
  \bibinfo{person}{Mohamed~Wiem Mkaouer}, {and} \bibinfo{person}{Emily Hill}.}
  \bibinfo{year}{2020}\natexlab{}.
\newblock \showarticletitle{On the generation, structure, and semantics of
  grammar patterns in source code identifiers}.
\newblock \bibinfo{journal}{\emph{Journal of Systems and Software}}
  \bibinfo{volume}{170} (\bibinfo{year}{2020}), \bibinfo{pages}{110740}.
\newblock
\showISSN{0164-1212}
\urldef\tempurl%
\url{https://doi.org/10.1016/j.jss.2020.110740}
\showDOI{\tempurl}


\bibitem[\protect\citeauthoryear{Petroni, Rocktäschel, Lewis, Bakhtin, Wu,
  Miller, and Riedel}{Petroni et~al\mbox{.}}{2019}]%
        {lmknowledge}
\bibfield{author}{\bibinfo{person}{Fabio Petroni}, \bibinfo{person}{Tim
  Rocktäschel}, \bibinfo{person}{Patrick Lewis}, \bibinfo{person}{Anton
  Bakhtin}, \bibinfo{person}{Yuxiang Wu}, \bibinfo{person}{Alexander~H.
  Miller}, {and} \bibinfo{person}{Sebastian Riedel}.}
  \bibinfo{year}{2019}\natexlab{}.
\newblock \showarticletitle{Language Models as Knowledge Bases?}. In
  \bibinfo{booktitle}{\emph{Proceedings of the 2019 Conference on Empirical
  Methods in Natural Language Processing and the 9th International Joint
  Conference on Natural Language Processing (EMNLP-IJCNLP)}}.
\newblock


\bibitem[\protect\citeauthoryear{Pradel and Sen}{Pradel and Sen}{2018}]%
        {cv5}
\bibfield{author}{\bibinfo{person}{Michael Pradel} {and}
  \bibinfo{person}{Koushik Sen}.} \bibinfo{year}{2018}\natexlab{}.
\newblock \showarticletitle{DeepBugs: A Learning Approach to Name-Based Bug
  Detection}.
\newblock \bibinfo{journal}{\emph{Proc. ACM Program. Lang.}}
  \bibinfo{volume}{2}, \bibinfo{number}{OOPSLA}, Article
  \bibinfo{articleno}{147} (\bibinfo{date}{Oct.} \bibinfo{year}{2018}),
  \bibinfo{numpages}{25}~pages.
\newblock
\urldef\tempurl%
\url{https://doi.org/10.1145/3276517}
\showDOI{\tempurl}


\bibitem[\protect\citeauthoryear{Rabin, Bui, Wang, Yu, Jiang, and
  Alipour}{Rabin et~al\mbox{.}}{2021}]%
        {new-2}
\bibfield{author}{\bibinfo{person}{Md~Rafiqul~Islam Rabin},
  \bibinfo{person}{Nghi~DQ Bui}, \bibinfo{person}{Ke Wang},
  \bibinfo{person}{Yijun Yu}, \bibinfo{person}{Lingxiao Jiang}, {and}
  \bibinfo{person}{Mohammad~Amin Alipour}.} \bibinfo{year}{2021}\natexlab{}.
\newblock \showarticletitle{On the generalizability of Neural Program Models
  with respect to semantic-preserving program transformations}.
\newblock \bibinfo{journal}{\emph{Information and Software Technology}}
  \bibinfo{volume}{135} (\bibinfo{year}{2021}), \bibinfo{pages}{106552}.
\newblock


\bibitem[\protect\citeauthoryear{Rubenstein and Goodenough}{Rubenstein and
  Goodenough}{1965}]%
        {distributional}
\bibfield{author}{\bibinfo{person}{Herbert Rubenstein} {and}
  \bibinfo{person}{John~B. Goodenough}.} \bibinfo{year}{1965}\natexlab{}.
\newblock \showarticletitle{Contextual Correlates of Synonymy}.
\newblock \bibinfo{journal}{\emph{Commun. ACM}} \bibinfo{volume}{8},
  \bibinfo{number}{10} (\bibinfo{date}{Oct.} \bibinfo{year}{1965}),
  \bibinfo{pages}{627–633}.
\newblock
\showISSN{0001-0782}
\urldef\tempurl%
\url{https://doi.org/10.1145/365628.365657}
\showDOI{\tempurl}


\bibitem[\protect\citeauthoryear{Schankin, Berger, Holt, Hofmeister, Riedel,
  and Beigl}{Schankin et~al\mbox{.}}{2018}]%
        {idf1}
\bibfield{author}{\bibinfo{person}{Andrea Schankin}, \bibinfo{person}{Annika
  Berger}, \bibinfo{person}{Daniel~V. Holt}, \bibinfo{person}{Johannes~C.
  Hofmeister}, \bibinfo{person}{Till Riedel}, {and} \bibinfo{person}{Michael
  Beigl}.} \bibinfo{year}{2018}\natexlab{}.
\newblock \showarticletitle{Descriptive Compound Identifier Names Improve
  Source Code Comprehension}. In \bibinfo{booktitle}{\emph{2018 IEEE/ACM 26th
  International Conference on Program Comprehension (ICPC)}}.
  \bibinfo{pages}{31--3109}.
\newblock


\bibitem[\protect\citeauthoryear{Shuai, Xu, Liu, Yan, Xia, and Lei}{Shuai
  et~al\mbox{.}}{2020}]%
        {shuai}
\bibfield{author}{\bibinfo{person}{Jianhang Shuai}, \bibinfo{person}{Ling Xu},
  \bibinfo{person}{Chao Liu}, \bibinfo{person}{Meng Yan}, \bibinfo{person}{Xin
  Xia}, {and} \bibinfo{person}{Yan Lei}.} \bibinfo{year}{2020}\natexlab{}.
\newblock \bibinfo{booktitle}{\emph{Improving Code Search with Co-Attentive
  Representation Learning}}.
\newblock \bibinfo{publisher}{Association for Computing Machinery},
  \bibinfo{address}{New York, NY, USA}, \bibinfo{pages}{196–207}.
\newblock
\showISBNx{9781450379588}
\urldef\tempurl%
\url{https://doi.org/10.1145/3387904.3389269}
\showURL{%
\tempurl}


\bibitem[\protect\citeauthoryear{Sinha, Jia, Hupkes, Pineau, Williams, and
  Kiela}{Sinha et~al\mbox{.}}{2021}]%
        {distributional2}
\bibfield{author}{\bibinfo{person}{Koustuv Sinha}, \bibinfo{person}{Robin Jia},
  \bibinfo{person}{Dieuwke Hupkes}, \bibinfo{person}{Joelle Pineau},
  \bibinfo{person}{Adina Williams}, {and} \bibinfo{person}{Douwe Kiela}.}
  \bibinfo{year}{2021}\natexlab{}.
\newblock \showarticletitle{Masked Language Modeling and the Distributional
  Hypothesis: Order Word Matters Pre-training for Little}. In
  \bibinfo{booktitle}{\emph{Proceedings of the 2021 Conference on Empirical
  Methods in Natural Language Processing}}.
\newblock


\bibitem[\protect\citeauthoryear{Svajlenko, Islam, Keivanloo, Roy, and
  Mia}{Svajlenko et~al\mbox{.}}{2014}]%
        {bigclone}
\bibfield{author}{\bibinfo{person}{Jeffrey Svajlenko},
  \bibinfo{person}{Judith~F. Islam}, \bibinfo{person}{Iman Keivanloo},
  \bibinfo{person}{Chanchal~K. Roy}, {and} \bibinfo{person}{Mohammad~Mamun
  Mia}.} \bibinfo{year}{2014}\natexlab{}.
\newblock \showarticletitle{Towards a Big Data Curated Benchmark of
  Inter-project Code Clones}. In \bibinfo{booktitle}{\emph{2014 IEEE
  International Conference on Software Maintenance and Evolution}}.
  \bibinfo{pages}{476--480}.
\newblock
\urldef\tempurl%
\url{https://doi.org/10.1109/ICSME.2014.77}
\showDOI{\tempurl}


\bibitem[\protect\citeauthoryear{Tenney, Das, and Pavlick}{Tenney
  et~al\mbox{.}}{2019a}]%
        {nlppipeline2}
\bibfield{author}{\bibinfo{person}{Ian Tenney}, \bibinfo{person}{Dipanjan Das},
  {and} \bibinfo{person}{Ellie Pavlick}.} \bibinfo{year}{2019}\natexlab{a}.
\newblock \showarticletitle{BERT rediscovers the classical NLP pipeline}. In
  \bibinfo{booktitle}{\emph{Proceedings of the 57th Annual Meeting of the
  Association for Computational Linguistics}}.
\newblock


\bibitem[\protect\citeauthoryear{Tenney, Xia, Chen, Wang, Poliak, McCoy, Kim,
  Durme, Bowman, Das, and Pavlick}{Tenney et~al\mbox{.}}{2019b}]%
        {nlpclass}
\bibfield{author}{\bibinfo{person}{Ian Tenney}, \bibinfo{person}{Patrick Xia},
  \bibinfo{person}{Berlin Chen}, \bibinfo{person}{Alex Wang},
  \bibinfo{person}{Adam Poliak}, \bibinfo{person}{R~Thomas McCoy},
  \bibinfo{person}{Najoung Kim}, \bibinfo{person}{Benjamin~Van Durme},
  \bibinfo{person}{Sam Bowman}, \bibinfo{person}{Dipanjan Das}, {and}
  \bibinfo{person}{Ellie Pavlick}.} \bibinfo{year}{2019}\natexlab{b}.
\newblock \showarticletitle{What do you learn from context? Probing for
  sentence structure in contextualized word representations}. In
  \bibinfo{booktitle}{\emph{International Conference on Learning
  Representations}}.
\newblock
\urldef\tempurl%
\url{https://openreview.net/forum?id=SJzSgnRcKX}
\showURL{%
\tempurl}


\bibitem[\protect\citeauthoryear{Vaswani, Shazeer, Parmar, Uszkoreit, Jones,
  Gomez, Kaiser, and Polosukhin}{Vaswani et~al\mbox{.}}{2017}]%
        {all-you-need}
\bibfield{author}{\bibinfo{person}{Ashish Vaswani}, \bibinfo{person}{Noam
  Shazeer}, \bibinfo{person}{Niki Parmar}, \bibinfo{person}{Jakob Uszkoreit},
  \bibinfo{person}{Llion Jones}, \bibinfo{person}{Aidan~N. Gomez},
  \bibinfo{person}{Lukasz Kaiser}, {and} \bibinfo{person}{Illia Polosukhin}.}
  \bibinfo{year}{2017}\natexlab{}.
\newblock \showarticletitle{Attention Is All You Need}. In
  \bibinfo{booktitle}{\emph{Advances in Neural Information Processing
  Systems}}.
\newblock


\bibitem[\protect\citeauthoryear{Vig}{Vig}{2019}]%
        {bertviz}
\bibfield{author}{\bibinfo{person}{Jesse Vig}.}
  \bibinfo{year}{2019}\natexlab{}.
\newblock \showarticletitle{A Multiscale Visualization of Attention in the
  Transformer Model}. In \bibinfo{booktitle}{\emph{Proceedings of the 57th
  Annual Meeting of the Association for Computational Linguistics: System
  Demonstrations}}. \bibinfo{publisher}{Association for Computational
  Linguistics}, \bibinfo{address}{Florence, Italy}, \bibinfo{pages}{37--42}.
\newblock
\urldef\tempurl%
\url{https://doi.org/10.18653/v1/P19-3007}
\showDOI{\tempurl}


\bibitem[\protect\citeauthoryear{Wan, Zhao, Yang, Xu, Ying, Wu, and Yu}{Wan
  et~al\mbox{.}}{2018}]%
        {ase}
\bibfield{author}{\bibinfo{person}{Yao Wan}, \bibinfo{person}{Zhou Zhao},
  \bibinfo{person}{Min Yang}, \bibinfo{person}{Guandong Xu},
  \bibinfo{person}{Haochao Ying}, \bibinfo{person}{Jian Wu}, {and}
  \bibinfo{person}{Philip~S. Yu}.} \bibinfo{year}{2018}\natexlab{}.
\newblock \showarticletitle{Improving Automatic Source Code Summarization via
  Deep Reinforcement Learning}. In \bibinfo{booktitle}{\emph{Proceedings of the
  33rd ACM/IEEE International Conference on Automated Software Engineering}}
  (Montpellier, France) \emph{(\bibinfo{series}{ASE 2018})}.
  \bibinfo{publisher}{Association for Computing Machinery},
  \bibinfo{address}{New York, NY, USA}, \bibinfo{pages}{397–407}.
\newblock
\showISBNx{9781450359375}
\urldef\tempurl%
\url{https://doi.org/10.1145/3238147.3238206}
\showDOI{\tempurl}


\bibitem[\protect\citeauthoryear{Wang, Zhang, Zeng, and Xu}{Wang
  et~al\mbox{.}}{2020}]%
        {trans3}
\bibfield{author}{\bibinfo{person}{Wenhua Wang}, \bibinfo{person}{Yuqun Zhang},
  \bibinfo{person}{Zhengran Zeng}, {and} \bibinfo{person}{Guandong Xu}.}
  \bibinfo{year}{2020}\natexlab{}.
\newblock \showarticletitle{TranS\^3: A Transformer-based Framework for
  Unifying Code Summarization and Code Search}.
\newblock \bibinfo{journal}{\emph{arXiv e-prints}}, Article
  \bibinfo{articleno}{arXiv:2003.03238} (\bibinfo{date}{March}
  \bibinfo{year}{2020}), \bibinfo{numpages}{arXiv:2003.03238}~pages.
\newblock
\showeprint[arxiv]{2003.03238}~[cs.SE]


\bibitem[\protect\citeauthoryear{White, Tufano, Martínez, Monperrus, and
  Poshyvanyk}{White et~al\mbox{.}}{2019}]%
        {cv3}
\bibfield{author}{\bibinfo{person}{Martin White}, \bibinfo{person}{Michele
  Tufano}, \bibinfo{person}{Matías Martínez}, \bibinfo{person}{Martin
  Monperrus}, {and} \bibinfo{person}{Denys Poshyvanyk}.}
  \bibinfo{year}{2019}\natexlab{}.
\newblock \showarticletitle{Sorting and Transforming Program Repair Ingredients
  via Deep Learning Code Similarities}. In \bibinfo{booktitle}{\emph{2019 IEEE
  26th International Conference on Software Analysis, Evolution and
  Reengineering (SANER)}}. \bibinfo{pages}{479--490}.
\newblock
\urldef\tempurl%
\url{https://doi.org/10.1109/SANER.2019.8668043}
\showDOI{\tempurl}


\bibitem[\protect\citeauthoryear{Wolf, Debut, Sanh, Chaumond, Delangue, Moi,
  Cistac, Rault, Louf, Funtowicz, Davison, Shleifer, von Platen, Ma, Jernite,
  Plu, Xu, Scao, Gugger, Drame, Lhoest, and Rush}{Wolf et~al\mbox{.}}{2020}]%
        {codeberta}
\bibfield{author}{\bibinfo{person}{Thomas Wolf}, \bibinfo{person}{Lysandre
  Debut}, \bibinfo{person}{Victor Sanh}, \bibinfo{person}{Julien Chaumond},
  \bibinfo{person}{Clement Delangue}, \bibinfo{person}{Anthony Moi},
  \bibinfo{person}{Pierric Cistac}, \bibinfo{person}{Tim Rault},
  \bibinfo{person}{Rémi Louf}, \bibinfo{person}{Morgan Funtowicz},
  \bibinfo{person}{Joe Davison}, \bibinfo{person}{Sam Shleifer},
  \bibinfo{person}{Patrick von Platen}, \bibinfo{person}{Clara Ma},
  \bibinfo{person}{Yacine Jernite}, \bibinfo{person}{Julien Plu},
  \bibinfo{person}{Canwen Xu}, \bibinfo{person}{Teven~Le Scao},
  \bibinfo{person}{Sylvain Gugger}, \bibinfo{person}{Mariama Drame},
  \bibinfo{person}{Quentin Lhoest}, {and} \bibinfo{person}{Alexander~M. Rush}.}
  \bibinfo{year}{2020}\natexlab{}.
\newblock \showarticletitle{HuggingFace's Transformers: State-of-the-art
  Natural Language Processing}. In \bibinfo{booktitle}{\emph{Proceedings of the
  2020 Conference on Empirical Methods in Natural Language Processing: System
  Demonstrations}}.
\newblock


\bibitem[\protect\citeauthoryear{Yang, Shi, He, and Lo}{Yang
  et~al\mbox{.}}{2022}]%
        {new-4}
\bibfield{author}{\bibinfo{person}{Zhou Yang}, \bibinfo{person}{Jieke Shi},
  \bibinfo{person}{Junda He}, {and} \bibinfo{person}{David Lo}.}
  \bibinfo{year}{2022}\natexlab{}.
\newblock \showarticletitle{Natural Attack for Pre-trained Models of Code}. In
  \bibinfo{booktitle}{\emph{2022 IEEE/ACM 41st International Conference on
  Software Engineering (ICSE)}}.
\newblock


\bibitem[\protect\citeauthoryear{Zhang, Wang, Zhang, Sun, and Liu}{Zhang
  et~al\mbox{.}}{2020}]%
        {rencos}
\bibfield{author}{\bibinfo{person}{Jian Zhang}, \bibinfo{person}{Xu Wang},
  \bibinfo{person}{Hongyu Zhang}, \bibinfo{person}{Hailong Sun}, {and}
  \bibinfo{person}{Xudong Liu}.} \bibinfo{year}{2020}\natexlab{}.
\newblock \showarticletitle{Retrieval-Based Neural Source Code Summarization}.
  In \bibinfo{booktitle}{\emph{Proceedings of the ACM/IEEE 42nd International
  Conference on Software Engineering}} (Seoul, South Korea)
  \emph{(\bibinfo{series}{ICSE '20})}. \bibinfo{publisher}{Association for
  Computing Machinery}, \bibinfo{address}{New York, NY, USA},
  \bibinfo{pages}{1385–1397}.
\newblock
\showISBNx{9781450371216}
\urldef\tempurl%
\url{https://doi.org/10.1145/3377811.3380383}
\showDOI{\tempurl}


\end{thebibliography}

%








\end{document}